\documentclass[aps,twocolumn,superscriptaddress,preprintnumbers,showpacs,amsmath,amssymb,nofootinbib]{revtex4}

\usepackage{graphicx} 
\usepackage{dcolumn}  
\usepackage{color}

\renewcommand{\arraystretch}{1.2}
\usepackage{epsfig}
\usepackage{epsbox}
\usepackage[T1]{fontenc}
\usepackage{wrapfloat}
\usepackage{subfigure}
\usepackage{here}
\usepackage{array}
\usepackage{amsmath, amssymb, hhline, multirow, hangcaption, subfigure}
\usepackage{graphics}
\usepackage{graphicx}
\usepackage{lscape}
\usepackage{longtable}
\usepackage{supertabular}
\usepackage{refmerge}

\def \vec #1{\mbox{{\boldmath $#1$}}}

\def \lr #1{\left( #1 \right)}

\def \B {{\cal B}}

\def \GeV {{\rm GeV}}
\def \MeV {{\rm MeV}}

\def \eV {{\rm eV}}

\begin{document}
\hspace{150mm}
\preprint{\vbox{ \hbox{   }
                 \hbox{Belle Preprint 2010-13}
                 \hbox{KEK Preprint 2010-20}
                 \hbox{July 2010}
                 \hbox{ (Rev. Dec. 2010) }
}}

\title
{\ \\Measurement of ${\boldmath \eta \eta}$ production
in two-photon collisions}

\begin{abstract}
We report the first measurement of 
the differential cross section for the process $\gamma \gamma \to \eta \eta$
in the kinematic range above the $\eta\eta$
threshold, $1.096~\GeV < W < 3.8~\GeV$ over nearly the entire solid angle range, 
$|\cos \theta^*| \le 0.9$ or $\le 1.0$ depending on $W$,
where $W$ and $\theta^*$ are the energy and $\eta$ scattering 
angle, respectively, in the $\gamma\gamma$ center-of-mass system.
The results are based on a 393~fb$^{-1}$ data sample
collected with the Belle detector at the KEKB $e^+ e^-$ collider.
In the $W$ range 1.1--2.0~GeV/$c^2$ we perform an analysis of resonance
amplitudes for various partial waves, and at higher energy we 
compare the energy and the angular dependences of the cross 
section with predictions of theoretical models and extract
contributions of the $\chi_{cJ}$ charmonia. 
\end{abstract}

\normalsize

\affiliation{Budker Institute of Nuclear Physics, Novosibirsk}
\affiliation{Faculty of Mathematics and Physics, Charles University, Prague}
\affiliation{Chiba University, Chiba}
\affiliation{University of Cincinnati, Cincinnati, Ohio 45221}
\affiliation{Justus-Liebig-Universit\"at Gie\ss{}en, Gie\ss{}en}
\affiliation{Hanyang University, Seoul}
\affiliation{University of Hawaii, Honolulu, Hawaii 96822}
\affiliation{High Energy Accelerator Research Organization (KEK), Tsukuba}
\affiliation{Institute of High Energy Physics, Chinese Academy of Sciences, Beijing}
\affiliation{Institute of High Energy Physics, Vienna}
\affiliation{Institute of High Energy Physics, Protvino}
\affiliation{Institute for Theoretical and Experimental Physics, Moscow}
\affiliation{J. Stefan Institute, Ljubljana}
\affiliation{Kanagawa University, Yokohama}
\affiliation{Institut f\"ur Experimentelle Kernphysik, Karlsruher Institut f\"ur Technologie, Karlsruhe}
\affiliation{Korea Institute of Science and Technology Information, Daejeon}
\affiliation{Korea University, Seoul}
\affiliation{Kyungpook National University, Taegu}
\affiliation{\'Ecole Polytechnique F\'ed\'erale de Lausanne (EPFL), Lausanne}
\affiliation{Faculty of Mathematics and Physics, University of Ljubljana, Ljubljana}
\affiliation{University of Maribor, Maribor}
\affiliation{Max-Planck-Institut f\"ur Physik, M\"unchen}
\affiliation{University of Melbourne, School of Physics, Victoria 3010}
\affiliation{Nagoya University, Nagoya}
\affiliation{Nara Women's University, Nara}
\affiliation{National Central University, Chung-li}
\affiliation{National United University, Miao Li}
\affiliation{Department of Physics, National Taiwan University, Taipei}
\affiliation{H. Niewodniczanski Institute of Nuclear Physics, Krakow}
\affiliation{Nippon Dental University, Niigata}
\affiliation{Niigata University, Niigata}
\affiliation{University of Nova Gorica, Nova Gorica}
\affiliation{Novosibirsk State University, Novosibirsk}
\affiliation{Osaka City University, Osaka}
\affiliation{University of Science and Technology of China, Hefei}
\affiliation{Seoul National University, Seoul}
\affiliation{Sungkyunkwan University, Suwon}
\affiliation{School of Physics, University of Sydney, NSW 2006}
\affiliation{Tata Institute of Fundamental Research, Mumbai}
\affiliation{Excellence Cluster Universe, Technische Universit\"at M\"unchen, Garching}
\affiliation{Toho University, Funabashi}
\affiliation{Tohoku Gakuin University, Tagajo}
\affiliation{Tohoku University, Sendai}
\affiliation{Department of Physics, University of Tokyo, Tokyo}
\affiliation{Tokyo Metropolitan University, Tokyo}
\affiliation{Tokyo University of Agriculture and Technology, Tokyo}
\affiliation{IPNAS, Virginia Polytechnic Institute and State University, Blacksburg, Virginia 24061}
\affiliation{Wayne State University, Detroit, Michigan 48202}
\affiliation{Yonsei University, Seoul}
\author{S.~Uehara}\affiliation{High Energy Accelerator Research Organization (KEK), Tsukuba} 
\author{Y.~Watanabe}\affiliation{Kanagawa University, Yokohama} 
\author{H.~Nakazawa}\affiliation{National Central University, Chung-li} 
  \author{I.~Adachi}\affiliation{High Energy Accelerator Research Organization (KEK), Tsukuba} 
  \author{H.~Aihara}\affiliation{Department of Physics, University of Tokyo, Tokyo} 
  \author{K.~Arinstein}\affiliation{Budker Institute of Nuclear Physics, Novosibirsk}\affiliation{Novosibirsk State University, Novosibirsk} 
  \author{T.~Aushev}\affiliation{\'Ecole Polytechnique F\'ed\'erale de Lausanne (EPFL), Lausanne}\affiliation{Institute for Theoretical and Experimental Physics, Moscow} 
  \author{A.~M.~Bakich}\affiliation{School of Physics, University of Sydney, NSW 2006} 
  \author{V.~Balagura}\affiliation{Institute for Theoretical and Experimental Physics, Moscow} 
  \author{E.~Barberio}\affiliation{University of Melbourne, School of Physics, Victoria 3010} 
% ...
\collaboration{The Belle Collaboration}

\renewcommand{\thefootnote}{\arabic{footnote}}

\setcounter{footnote}{0}
\setcounter{figure}{0}

\normalsize

\maketitle
\normalsize

\section{Introduction}
\label{sec-1}
Measurements of exclusive hadronic final states in two-photon
collisions provide valuable information concerning the physics of light and 
heavy-quark resonances, perturbative and non-perturbative QCD 
and hadron-production mechanisms.
So far, we, the Belle Collaboration,
have measured the production cross sections for 
charged-pion pairs~\cite{mori1,mori2,nkzw},
charged and neutral-kaon pairs~\cite{nkzw,kabe,wtchen},
and proton-antiproton pairs~\cite{kuo}.
We have also analyzed $D$-meson-pair production and observe a new
charmonium state identified as the $\chi_{c2}(2P)$~\cite{uehara}.
Recently, we have examined $\omega J/\psi$ and $\phi J/\psi$ production
and also found charmonium-like structures in these final 
states~\cite{uehara2,shen}.

In addition, we have measured the production cross section
for  the $\pi^0 \pi^0$ and $\eta \pi^0$ 
final states~\cite{pi0pi0,pi0pi02, etapi0}.
The statistics of these measurements 
are
 two to three orders of
magnitude higher than in pre-B-factory measurements~\cite{past_exp}, 
opening a new era in studies of two-photon physics.

In the present study, we report measurements of the differential 
cross sections,
$d\sigma/d|\cos \theta^*|$,
for the process $\gamma \gamma \to \eta \eta$ in
a wide two-photon center-of-mass (c.m.) energy ($W$) range 
from the $\eta\eta$ mass threshold 1.096~GeV to 3.8~GeV, and
in the c.m. angular range, $|\cos \theta^*| \leq 1$ ($0.9$)
for $W \leq 2.0~\GeV$ ($W > 2.0~\GeV$).
In this analysis, we use the
$\eta \to \gamma \gamma$ decay mode only because
the $\eta \to \pi^+\pi^-\pi^0$ decay mode has 
a much smaller product of efficiency and
branching fraction.

The $I^G J^{PC}$ quantum numbers
of a meson produced by two photons and decaying into $\eta \eta$ 
are restricted to be
$0^+$(even)$^{++}$, that is, those of $f_{J={\rm even}}$ 
or $\chi_{cJ={\rm even}}$ mesons.
A long-standing puzzle in QCD
is the existence and structure of low mass scalar mesons.
In the $I=0$
sector, we recently observed 
a peaking structure at the $f_0(980)$ mass 
in both
the $\gamma\gamma \to \pi^+\pi^-$ and $\gamma\gamma \to \pi^0\pi^0$ 
channels~\cite{mori1,pi0pi0}.
Our analysis also suggests the existence of another $f_0$ meson
in the 1.2--1.5~GeV region that couples to two photons~\cite{pi0pi0}.
The significant $s\bar{s}$ component in the $\eta$ meson implies
a connection of this reaction to the $K^+K^-$~\cite{kabe} 
and $K^0 \bar{K}^0$~\cite{wtchen} 
processes.

At higher energies ($W > 2.4$~GeV), we can invoke a quark model.
In leading-order calculations, the ratio of the $\eta \pi^0$ or 
$\eta \eta$ cross section 
to that of $\pi^0 \pi^0$ is predicted. 
Analyses of energy and angular distributions of these cross sections
are essential to determine properties of the observed
resonances and to test the validity of QCD based models~\cite{bl,handbag,
handbag2} involving $q \bar{q}$ production and SU(3) flavor symmetry.
It is also 
interesting
to compare the behavior of $\eta\eta$ production
with that of $K^+K^-$ and $K^0_S K^0_S$, which
have been measured by the Belle experiment~\cite{nkzw,wtchen}. 
The cross section for
the $\gamma \gamma \to \eta \eta$ process has not been measured so far.

The organization of this paper is as follows.
In Sec.~II, the experimental apparatus and event selection are described.
Signal yields and backgrounds are discussed in Sec.~III.
Differential cross sections are then extracted in Sec.~IV.
In Sec.~V the $f_2(1270)$, $f_2'(1525)$ and other possible resonances
are studied by parameterizing partial wave amplitudes.
The behavior of differential cross sections and $W$ dependence of 
the integrated cross sections at higher energy region ($W>2.4~\GeV$)
are compared to QCD predictions in Sec.~VI.
Finally in Sec.~VII, a summary and conclusion are given.

\section{Experimental apparatus and event selection}
\label{sec-2}
Events with all neutral final states are extracted from the data 
collected by the Belle experiment.
In this section, the Belle detector and event selection procedure are
described.

\subsection{Experimental apparatus}
A comprehensive description of the Belle detector is
given elsewhere~\cite{belle}.
We mention here only those
detector components that are essential for the present measurement.
Charged tracks are reconstructed from hit information in 
the silicon vertex detector and the central
drift chamber (CDC) located in a uniform 1.5~T solenoidal magnetic field.
The detector solenoid is oriented along the $z$ axis, which points
in the direction opposite to that of the positron beam. 
Photon detection and
energy measurements are performed with a CsI(Tl) electromagnetic
calorimeter (ECL).

For this all-neutral final state, we require that there be no
reconstructed tracks coming from the vicinity of
the nominal collision point. 
Therefore, the CDC is used to veto events with charged track(s). 
The photons from a decay of the $\eta$ meson
are detected and their momentum vectors are measured by the ECL. 
The ECL is also used to trigger signal events.
Two kinds of ECL triggers
are used to select events of interest: the total ECL energy deposit
in the 
acceptance region used by the trigger (see the next subsection)
is greater than 1.15~GeV (the ``HiE'' trigger), or
four or more
ECL clusters above an
energy threshold of 110~MeV in segments
of the ECL (the ``Clst4'' trigger).
The above energy thresholds are determined by studying the
correlations between the two triggers in the experimental data.

\subsection{Experimental data and data filtering}
We use a 393~fb$^{-1}$ data sample accumulated by the Belle detector 
at the KEKB asymmetric-energy $e^+e^-$ collider~\cite{kekb}.
For an early part of Belle data taking,
all neutral final states were not recorded.
Thus, this data set is smaller than the hadronic data
sample 
available at Belle.

The data were recorded at several $e^+e^-$ c.m. energy regions 
summarized in Table~\ref{tab:lum_data}.
We combine the results from the different beam energies,
because the $e^+e^-$ c.m. energy is more than
twice our $\gamma\gamma$ c.m. energy range for
any of the beam energies, and the beam-energy dependence of
the two-photon luminosity function is rather small. 
We generate most of the signal Monte-Carlo (MC) 
events and calculate the two-photon luminosity function
for 10.58~GeV. We then derive a correction factor 
for the other beam energies. 
The correction is less
than 0.5\% over the full range of $\gamma \gamma$ cm energies considered here.
The signal MC and the beam energy dependences are described in Sec.~IV.C.

The analysis is carried out in the ``zero-tag'' mode, where
neither the recoil electron nor positron are detected. 
We restrict the virtuality of the incident photons to be small
by imposing a strict requirement on the transverse-momentum balance 
with respect to the beam axis for the final-state hadronic system.

The filtering procedure (``Neutral Skim'') used for this analysis
is the same as the one used for
$\pi^0 \pi^0$ and $\eta \pi^0$ studies~\cite{pi0pi0,pi0pi02,etapi0}. 
The important requirements in this filter are the following:
there are no tracks originating in the beam collision region
and having a transverse momentum greater than 0.1~GeV/$c$
in the laboratory frame;
two or more photons that satisfy a specified energy or 
transverse-momentum criterion;
this requirement is
satisfied when there are three or more photons 
each with an energy above 100~MeV.
The performance of the ECL
triggers is studied in detail using $\pi^0 \pi^0$ events~\cite{pi0pi0}. 
We also study 
the trigger thresholds
using the $\eta \eta$ signal samples.

\begin{table*}
\caption{Data sample: integrated luminosities and energies}
\label{tab:lum_data}
\begin{center}
\begin{tabular}{ccl} \hline \hline
~~$e^+e^-$ c.m. energy~~ & Integrated luminosity & Runs \\
(GeV)& (fb$^{-1}$) & \\ \hline
10.58 & 286 & $\Upsilon(4S)$ \\ 
10.52 & 33 & continuum\\
9.43 - 9.46 & 7.3  & near  $\Upsilon(1S)$\\
9.99 - 10.03 & 6.7 &  near  $\Upsilon(2S)$\\
10.32 - 10.36 & 3.2 & near $\Upsilon(3S)$ \\
10.83 - 11.02 & 58 & near $\Upsilon(5S)$ \\ \hline
Total & 393 & \\
\hline\hline
\end{tabular}
\end{center}
\end{table*}

\subsection{Event selection}
From the Neutral Skim event sample,
we select $\gamma \gamma \to
\eta \eta$ candidates that satisfy the following conditions: 
\newcounter{num}
\begin{list}%
{(\arabic{num})}{\usecounter{num}}
\setlength{\rightmargin}{\leftmargin}
\setlength{\topsep}{0mm}
\setlength{\parskip}{0mm}
\setlength{\parsep}{0mm}
\setlength{\itemsep}{0mm}
\item the total energy deposit in ECL
is less
 than 5.7~GeV;
\item 
each photon candidate is required to have 
an energy of at least 100~MeV, and events
with four such photons are selected;
\item the event is triggered by either the ECL trigger HiE or Clst4;
\item 
either the sum of the energies of the  photons in  
the acceptance region used by the trigger  
is larger than 1.25~GeV, or all four selected photons 
are within this region, where the trigger acceptance is
the polar-angle range, $-0.6255 < \cos \theta <+0.9563$,
in the laboratory frame;

\item 
of the three possible combinations that can be constructed from
the four photons, there is one in which each invariant mass 
of the two photon pairs satisfies 
0.52~GeV/$c^2 < M_{\gamma\gamma\ i} <$ 0.57~GeV/$c^2$, where
 $i=1$, 2 is an index of the two-photon pairs;
\item
there is no neutral pion combination that is constructed
from any two of the four photons with a $\chi^2$ smaller than
9 in the mass-constrained fit;
\item the transverse momentum for the $\eta \eta$ system
$|\Sigma {\vec p}_t^*|$
is required to be less than $0.05~\GeV/c$.
\end{list}

A small fraction of events contain multiple combinations
of the four photons that satisfy criterion (5).
In those events, we take only one combination whose residual
for the nominal $\eta$ mass ($m_\eta = 0.5478$~GeV/$c^2$),
$(M_{\gamma\gamma 1} - m_\eta)^2 +(M_{\gamma\gamma 2} - m_\eta)^2 $,
is the smallest.
 
We then scale the energy of the two photons 
with a factor that is the ratio of the nominal $\eta$ mass to the 
reconstructed mass, $m_\eta/M_{\gamma\gamma\ i}$.
This is equivalent to an approximate 1C (one constraint) mass constraint 
fit in which the relative energy resolution ($\Delta E/E$) is independent of $E$
and the resolution in the angle measurement is much better than
that of the energy. 
This is a good approximation for
the $\eta$'s in this momentum range.
Using the corrected four-momenta of the $\eta$ mesons,
we calculate the invariant mass ($W$) and the transverse 
momentum ($|\Sigma {\vec p}_t^*|$)
in the $e^+e^-$ c.m. frame for the $\eta \eta$ system
and apply cut (7) above. 
We select 31655 candidates in the
region $W < 4.0$~GeV.

\begin{figure*}
\centering
\includegraphics[width=10cm]{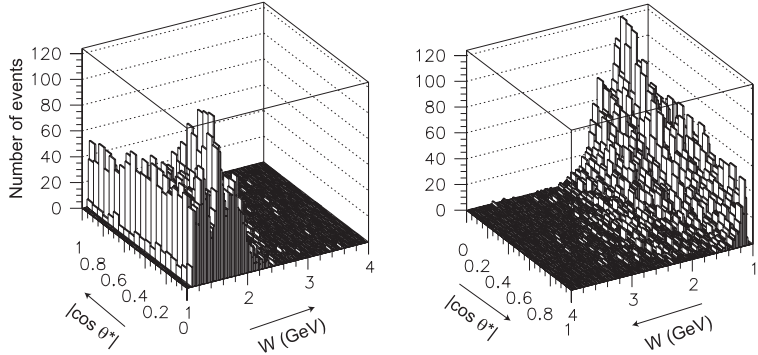}
\caption{Two-dimensional $W$ and $|\cos \theta^*|$ distribution 
for the $\eta \eta$ candidates in data.
The same distribution is viewed from two different directions.
The backgrounds are not subtracted.}
\label{fig01}
\end{figure*}

We define the c.m. scattering angle, $\theta^*$,
as the scattering angle of the $\eta$. 
The $e^+e^-$ direction is used 
to approximate
the axis for the polar angle calculation
because the exact $\gamma\gamma$ axis is unknown
for  untagged events.
The two-dimensional 
($W$, $|\cos \theta^*|$)
distribution of selected events is shown in Fig.~\ref{fig01}.

The probability for a signal $\eta\eta$ event to have
multiple combinations is 
sizable only near the threshold (about 6\% at 
$W \sim 1.11$~GeV), but it is small (less than 2\%) 
above $W>1.12$~GeV, according to the signal MC samples.
For different choices of $\gamma$-pair combinations in an event, 
the $W$ values are nearly the same,
but $|\cos \theta^*|$ can be 
different. 
As the angular distribution is observed to be flat
near the threshold, which is also theoretically expected,
the effect of an incorrect choice is negligibly small.

\section{Yields of the signal and backgrounds}
\label{sec-3}
In this section, backgrounds are identified and subtracted and
the extraction of the signal yield is discussed.

\subsection{Determination of non-{\boldmath $\eta \eta$} background}
There are two kinds of background processes for the 
$\gamma \gamma \to \eta \eta$ signal process: non-$\eta\eta$ and
$\eta\eta X$ backgrounds.
The non-$\eta\eta$ background does
not contain an $\eta$ pair in the final state, while the
$\eta\eta X$ background includes
extra particle(s) in the final state in addition to 
the $\eta\eta$ combination. In this 
measurement, the non-$\eta \eta$ contribution, 
arising from beam backgrounds or other physics
processes, is the dominant background in the final sample.

We first determine the number of the non-$\eta\eta$ background events 
using the yields of $\eta\eta$ mass sidebands. After subtracting 
this background contribution,
we check the $p_t$-balance distribution for the remaining component;
the signal component peaks near $|\Sigma {\vec p}_t^*|=0$, while
we expect that the $\eta\eta X$ background does not.  

\subsubsection{Defining the $\eta\eta$-mass sidebands}
\begin{figure*}
\centering
\includegraphics[width=10cm]{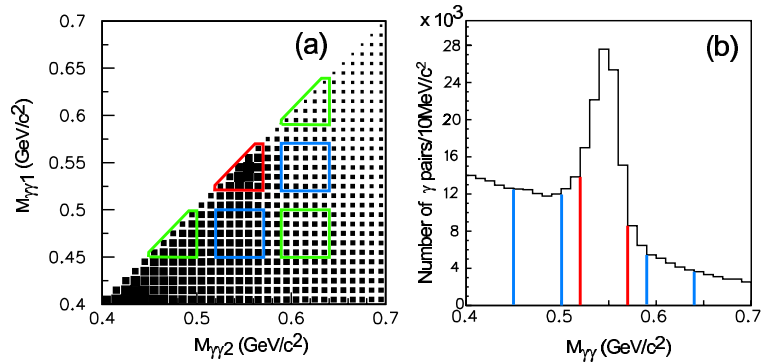}
\caption{(a) Two-dimensional $M_{\gamma\gamma 1}$ vs.
$M_{\gamma\gamma 2}$ distribution 
of the four-photon 
candidates
in data. 
We take  $M_{\gamma\gamma 1} < M_{\gamma\gamma 2}$.
The $p_t$-balance cut with $p_t < 0.1~$GeV/$c$ is applied using the 
photon momenta before the $m_\eta$ mass correction, which
reduces backgrounds. 
Red, blue and green boxes show the
signal, Sideband A and Sideband B regions, respectively. (b) A
one-dimensional projection of the same distribution where the $\gamma \gamma$
pair on the opposite side is required to be in the signal mass region, 0.52--0.57~GeV/$c^2$.
The vertical red and blue lines show the signal and two Sideband A regions,
respectively. Note that there are two entries per event in the
signal region.}
\label{fig02}
\end{figure*}

The $\eta\eta$-mass sidebands are defined by displacing
the central points of the mass intervals in selection criterion (5) 
by $\pm 0.07$~GeV/$c^2$.
Two kinds of sidebands are defined: Sideband A and Sideband B.
In Sideband A, the central points for the two-dimensional
mass cut for ($M_{\gamma \gamma 1}$, $M_{\gamma \gamma 2}$) are
(here, we assume $M_{\gamma \gamma 1} < M_{\gamma \gamma 2}$)
(0.545, 0.615) and (0.475, 0.545) in units of GeV/$c^2$ and
the width of the range is $\pm 0.025$~GeV/$c^2$. 
Sideband B has central points (0.475, 0.475), (0.475, 0.615)
and  (0.615, 0.615).  
When there are two or more choices of $\gamma$-pair  combinations
in an event that fall in the same sideband box, we take
the one  
that is closest
to the nominal central point of
each sideband box, ($m_\eta - 0.07$, $m_\eta$) or 
($m_\eta$, $m_\eta+0.07$) for Sideband A, and  
($m_\eta +/- 0.07$, $m_\eta +/- 0.07$) for Sideband B.
This is similar to the multiple candidate
selection applied
for the signal candidates.
The $M_{\gamma \gamma}$ distributions near the signal 
and sideband
regions are shown in Fig.~\ref{fig02}.

We also calculate $W$ and $|\cos \theta^*|$  
for the Sideband A and B candidates by scaling 
$M_{\gamma \gamma\ i}$ to $m_\eta$ (not to $m_\eta +/- 0.07$,
which would change the threshold mass).

\subsubsection{Sideband subtraction}
\begin{figure}
\centering
\includegraphics[width=6cm]{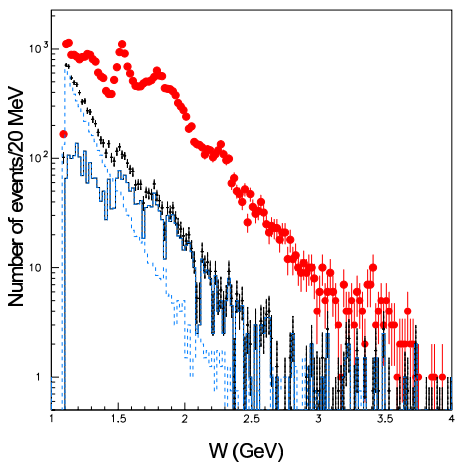}
\caption{ $W$ distributions of the yields in the signal region
(closed circles
with error bars, $Y_s$), and estimates of backgrounds
of the $\eta$~non-$\eta$ component (solid histogram, 
$0.5(Y_{bA}-Y_{bB})$) and the non-$\eta$~non-$\eta$ component 
(dashed histogram, $0.25Y_{bB}$).
Points
with error bars show the estimated total background,
$0.5 Y_{bA} - 0.25 Y_{bB}$.}

\label{fig03}
\end{figure}

We subtract the sideband yield to obtain the signal
component with the following formula:
\[
Y = Y_s - 0.5 Y_{bA} + 0.25 Y_{bB},
\]
where $Y$ is the signal yield after sideband
subtraction, $Y_s$ is the yield in the observed events
in the signal region, and
$Y_{bA}$ ( $Y_{bB}$ ) the yield of the Sideband A (B)
region. 
Here we model the non-$\eta$ backgrounds
with a linear distribution in $M_{\gamma \gamma}$,
for backgrounds with both $\eta$ non-$\eta$ 
and non-$\eta$ non-$\eta$ combinations.
The possibility of a non-linear background
component is included in the systematic
error (see Sec.~IV.E).
The yield in the signal and sideband regions 
(before the sideband subtraction) is
shown in Fig.~\ref{fig03}.
To obtain the differential cross sections, 
we subtract bin-by-bin
in each two-dimensional bin of $(W, |\cos \theta^*|)$
with bin widths  $\Delta W = 20~\MeV$ 
and  $\Delta |\cos \theta^*| = 0.1$. 
Two or five $W$ bins are combined later in 
the determination of the final cross sections.
Signal leakage into the sideband regions, which
amounts to 2--5\% of the signal size $Y$ 
and is larger at small $W$, 
is expected according to signal MC simulations. 
This effect is corrected in the derivation of the differential
cross sections by 
reducing
the efficiency.

\subsection{$p_t$-unbalanced component}
\begin{figure}
\centering
\includegraphics[width=6cm]{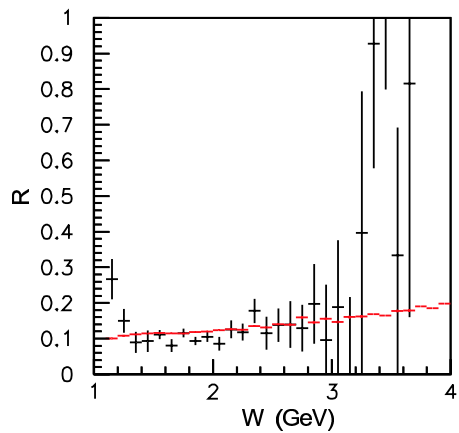}
\centering
\caption{Energy dependence of $R$ defined in Eq.(\ref{eqn:r_pt}).
It indicates the
level of $\eta \eta X$ background contamination, for the experimental data
(after the sideband subtraction,  
points with error bars) 
and signal MC (short horizontal bars).
}
\label{fig04}
\end{figure}

We expect that the background remaining
after the $\eta \eta$ sideband subtraction is very small.
To confirm this, we examine the $W$ dependence of the yield ratio 
of the $p_t$-unbalanced component
$R$ defined as:
\begin{equation}
R= \frac{Y(0.15~\GeV/c < |\sum \vec{p}_t^*| < 0.20~\GeV/c)}
  {Y (|\sum \vec{p}_t^*|<0.05~ \GeV/c)} \; ,
\label{eqn:r_pt}
\end{equation}
where $Y$ is the yield after the sideband subtraction 
in the specified $p_t$ region.
$R$ is plotted as a function of $W$ in Fig.~\ref{fig04},
where any excess over the signal MC would indicate
a contribution from $\eta\eta X$ background.
There is such a small excess just above mass threshold.
We include the effect from this possible background
source into
the correction and 
the systematic error.
Non-$\eta\eta$ background is
much larger near the $W$ threshold,
and this excess may be due to an imperfect sideband subtraction.
We apply a $-3\%$ correction
for the background from this source for $W<1.2$~GeV.
In the $W$ region between 1.2~GeV and 3.3~GeV, the $R$ value is 
consistent with the signal MC simulation. 
The reason why
the experimental data seems to be slightly below the MC 
for $R$ in the 1.4 -- 2.0~GeV range is not known, but the difference
translated to the background ratio is negligibly small, less than 1\%. 
For $W > 3.3$~GeV, there could be much larger 
$\eta \eta X$ backgrounds. 
As described in Sec.~IV.C, we do not report cross section
results for $W > 3.3$~GeV,
and apply a $-3\%$ correction for 3.2~GeV$< W <3.3$~GeV.

We conclude that this kind of background is
less than 2\% throughout the $W$ region, 1.2 -- 3.2~GeV, 
and assign 2\% as the systematic error for this source 
for the entire $W$ region, 1.096 -- 3.3~GeV.
These factors are 
obtained by assuming a quasi-linear $|\sum \vec{p}_t^*|$
dependence of the background and extracting its leakage into the
signal region ($|\sum \vec{p}_t^*|<0.05~\GeV/c$), which is approximately
1/6 of the
yield in the $0.15~\GeV/c < |\sum \vec{p}_t^*|<0.20~\GeV/c$ region.

\section{Deriving differential cross sections}
\label{sec-4}
In this section, we present the procedure to derive differential cross 
sections. 

\subsection{Effect of $e^+e^-$ beam energy }
We generate standard MC events for an $e^+e^-$ c.m. energy 
$\sqrt{s}=10.58$~GeV. 
We compare the products of the luminosity function
and efficiency  ($L_{\gamma\gamma}(W) \epsilon$ in Eq.~(2))
at three different c.m. energies, 9.46~GeV($\Upsilon(1S)$),
10.58~GeV($\Upsilon(4S)$) and 10.87~GeV($\Upsilon(5S)$) using 
MC samples. 
We conclude that, taking into account the integrated luminosities of
the different c.m. energies, the correction factors for the lower and
higher energy samples cancel almost exactly. 
Applying the MC results for
10.58 GeV to all samples leads to negligibly small effects of less than
0.5\%.

\subsection{Invariant mass resolution}
We estimate the invariant mass resolution of
the $\eta \eta$ system using the signal MC simulation.
Since we apply an energy rescaling using the $\eta$ mass,
the $W$ resolution is better than that for a pure 
energy measurement. 
We find that the invariant mass resolution is about 0.6\%
near the threshold, $W=1.1$--1.5~GeV and approaches
1.0\% for higher $W$. 
We confirm that the 
experimental resolution is at most 10\% larger 
than the MC resolution from measurements of
$p_t$ balance in $\pi^0 \pi^0$ production and in
the $\eta'$ peak in 
$\gamma \gamma \to \eta' \to \gamma \gamma$.
The resolution is much smaller than the $W$ bin 
widths: $\Delta W=0.04$~GeV or 0.1~GeV. 
Since statistics are low, we do not 
unfold our results as in 
previous measurements~\cite{pi0pi0,pi0pi02,etapi0}.

\subsection{Determination of the efficiency}
The signal MC simulations for $e^+e^- \to e^+e^- \eta\eta$ are
generated using the TREPS code~\cite{treps} and are used for
the efficiency calculation at 32 fixed $W$ points between 1.1 and 4.0~GeV
and isotropically in $|\cos \theta^*|$. 
We evaluate the efficiencies separately in $|\cos \theta^*|$ bins
with a width of 0.05, and thus the angular distribution at the generator
level does not play a role in the efficiency determination.

The $Q^2_{\rm max}$ parameter
that gives a maximum virtuality of the incident
photons is set to 1.0~GeV$^2$, 
while the cross sections for
virtual photon collisions include a form factor,
$\sigma_{\gamma\gamma}(0,Q^2)=\sigma_{\gamma\gamma}(0,0)/
(1+Q^2/W^2)^2$. 
Our analysis is not sensitive to the
form factor assumption,
since
our stringent $p_t$-balance cut
($|\sum \vec{p}_t^*| < 0.05~\GeV/c$) implies $Q^2/W^2$ is much smaller 
than unity; an approximate relation $Q^2 \sim |\sum \vec{p}_t^*|^2$ holds
when only one incident photon is treated as moderately virtual and 
the scattering angle of an electron (or a positron) that has emitted 
the virtual photon is small. 
Using signal MC simulation 
and replacing the $Q^2/W^2$ term by either $Q^2/m_\rho^2$ or omitting it
entirely, we confirm that the effect of the form factor choice on 
the cross section 
is less than 0.5\%, where $m_\rho$ is the $\rho$ meson mass.

Samples of 400,000 events are
generated at each $W$ point and are passed through 
the detector and trigger simulations. 
The obtained efficiencies are fitted to a two-dimensional function
of ($W$, $|\cos \theta^*|$) with an empirical functional form.

\begin{figure}
\centering
\includegraphics[width=5cm]{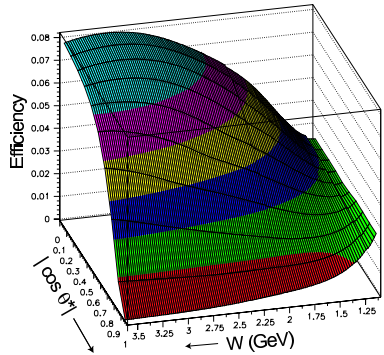}
\centering
\caption{ Two-dimensional dependence of the efficiency
on ($W$, $|\cos \theta^*|$).}
\label{fig05}
\end{figure}

We embed background hit patterns from random trigger data
into MC events. 
We find that different samples of background hits 
give small variations in the selection efficiency determination.
A $W$-dependent error in the efficiency, 3 -- 4\%, arises from 
the uncertainty in this effect.
Figure~\ref{fig05} shows the two-dimensional dependence 
of the efficiency on ( $W$, $|\cos \theta^*|$) 
after the smoothing fit. 

\subsection{Derivation of differential cross sections}
The differential cross section for each
($W$, $|\cos \theta^*|$) point is given by:
\begin{equation}
\frac{d\sigma}{d|\cos \theta^*|} =
\frac{\Delta Y }{\Delta W \Delta |\cos \theta^*| 
\int{\cal L}dt \ L_{\gamma\gamma}(W) \  \epsilon \ {\cal B}^2 }  ,
\label{eqn:dcs}
\end{equation}
where $\Delta Y$ is the signal yield after the $\eta$-mass
sideband subtraction, 
$\Delta W$ and $\Delta |\cos \theta^*|$ are the bin widths, 
$\int{\cal L}dt$ and  $L_{\gamma\gamma}(W)$ are
the integrated luminosity and two-photon luminosity function
calculated with TREPS~\cite{treps}, respectively,  $\epsilon$ is the  
efficiency, and ${\cal B}^2$ is the squared
branching fraction for $\eta \to \gamma \gamma$.
The over-subtraction of signal
in the sideband due to the leakage of the signal into
the sideband region is evaluated in the MC, separately, and
finally included in the efficiency $\epsilon$.

The bin sizes $\Delta W$ and $\Delta |\cos \theta^*|$ and the maximum
$|\cos \theta^*|$ for which we obtain the differential cross
section are summarized in
Table~\ref{tab:binsize}. We first derive the differential
cross sections for bin widths of $\Delta W = 0.02$~GeV and
 $\Delta |\cos \theta^*| = 0.1$, and average the differential
cross section over two or five different $W$ regions to obtain
results for $\Delta W = 0.04$~GeV or 0.10~GeV, respectively.

We do not give a cross section for $W>3.3$~GeV. In the $W$
range 3.3--3.6~GeV,
the charmonium component dominates the yield, and we cannot subtract
it in a model-independent way. We also cannot give the cross
section including the charmonium contribution in these bins,
because leakages from
the narrow $\chi_{c0}$ peak around 3.41~GeV
into adjacent bins
due to energy resolution
complicate the extraction of cross sections in each bin.
Above $W>3.6$~GeV, we do not find any significant 
signal after consideration of the backgrounds.

\begin{figure*}
\centering
\includegraphics[width=12cm]{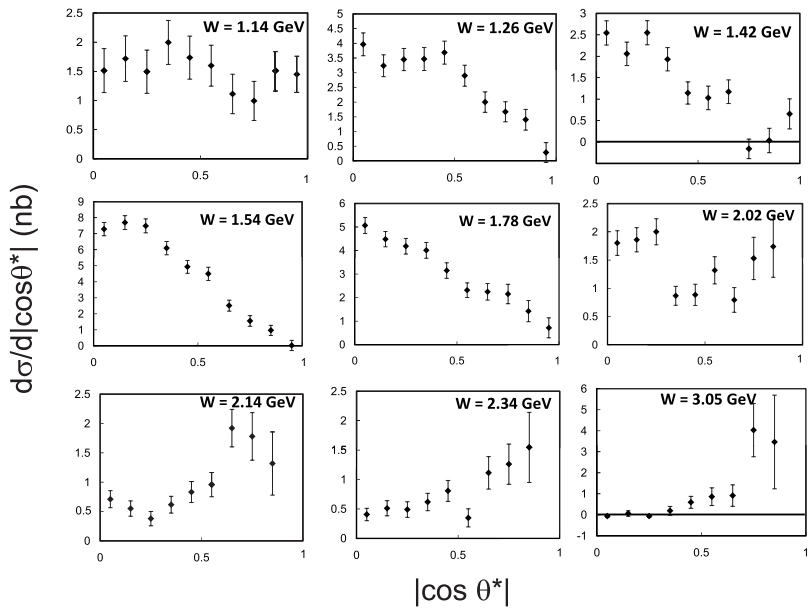}
\centering
\caption{Angular dependence of the 
differential cross sections for 
nine selected $W$ bins indicated.
The bin sizes are summarized in Table~\ref{tab:binsize}.} 
\label{fig06}
\end{figure*}

\begin{figure}
\centering
\includegraphics[width=6.cm]{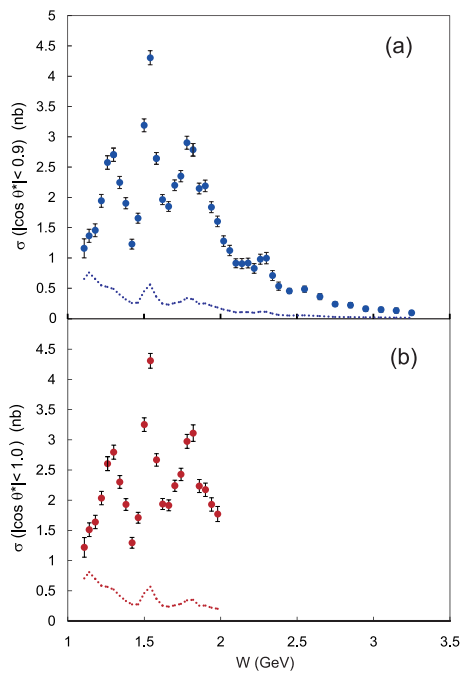}
\centering
\caption{
(a) The cross section integrated 
over $|\cos \theta^*|< 0.9$ and (b) over
$|\cos \theta^*|< 1.0$ for $W < 2.0$~GeV.
Errors are statistical only. The dotted curve
shows the size of the systematic uncertainty.} 
\label{fig07}
\end{figure}

Figure~\ref{fig06} shows the angular dependence of the 
differential cross sections for selected $W$ bins. 
Figure~\ref{fig07} shows the cross section integrated 
over $|\cos \theta^*|<0.9$ for the entire $W$ range and 
that for $|\cos \theta^*|<1.0$ in the range $W <2.0$~GeV.

\begin{table}
\caption{Bin sizes. 
The lowest bound of the first $W$ range
(1.0957~GeV) corresponds to the mass threshold.
}
\label{tab:binsize}
\begin{center}
\begin{tabular}{lccc} \hline \hline
$W$ range & $\Delta W$ & $\Delta |\cos \theta^*|$ & maximum\\ 
(GeV) & (GeV) & &  
$|\cos \theta^*|$ \\
\hline
1.0957 -- 1.12 & 0.0243 & 0.1 & 1.0 \\
1.12 -- 2.0 & 0.04 & 0.1 & 1.0 \\
2.0 -- 2.4 & 0.04 & 0.1 & 0.9 \\
2.4 -- 3.3 & 0.10 & 0.1 & 0.9 \\
\hline\hline
\end{tabular}
\end{center}
\end{table}

\subsection{Systematic errors}
Various sources of systematic uncertainties assigned
for the $\eta \eta$ signal yield, efficiency and the cross section
evaluation are described in detail below
and summarized in Table~\ref{tab:sysdcs}.
\setcounter{num}{0}
\begin{list}%
{(\arabic{num})}{\usecounter{num}}
\setlength{\rightmargin}{\leftmargin}
\setlength{\topsep}{0mm}
\setlength{\parskip}{0mm}
\setlength{\parsep}{0mm}
\setlength{\itemsep}{0mm}
\item {\it Trigger efficiency}: The systematic error due to  
uncertainty in 
the threshold for the Clst4 trigger ($\sim 110$~MeV) 
is very small, 
because photons from $\eta$ decays have high enough energy. 
However, the efficiency of the
HiE trigger dominates that of Clst4 except in the lowest $W$ region,
because the former has a looser condition for the number
of clusters in the 
acceptance region of the trigger.
We estimate the uncertainty in the efficiency for the 
HiE trigger to be 4\% over the whole $W$ region, and 
treat it as the combined
systematic error for the two kinds of triggers.

\item {\it 
$\eta$ selection
efficiency}: 
We assign 6\% for the 
selection
of the two $\eta$'s. This corresponds to a 3\% 
uncertainty for the efficiency of each $\eta$ 
reconstruction.
\item 
{\it Overlapping hits from beam background and related effects}: 
We assign a 4\% (3\%) 
error for $W < 1.44$~GeV ($W > 1.44$~GeV) 
for uncertainties of the inefficiency in event selection 
due to beam-background photons, which affect the photon multiplicity
and $\eta$ reconstruction. 
The uncertainty is estimated by comparing
efficiencies among different experimental periods 
and background conditions.
We adopt the average efficiency from
different background files, and the uncertainty in the average,
obtained from 
the variation of experimental yield in 
different run periods, 
is assigned as the error.

\item{\it $p_t$-balance cut}: 
A 3\% uncertainty is assigned. 
The $p_t$-balance distribution for the signal is well reproduced 
by MC so that the efficiency is correct to within this error.

\item {\it Sideband background subtraction}: 1/3 of the size of the 
subtracted component is assigned
to this source for each bin.
We conservatively assign this error because
we ignore the non-linear behavior of the background 
in the $M_{\gamma\gamma}$ distribution in the sideband
subtraction. This effect is expected to be large
but cannot be determined precisely in the
lowest $W$ bins.

\item 
{\it $p_t$-unbalanced background}:
We have applied a $-3\%$ correction for this
background source only in the lowest and highest
$W$ regions, $W<1.2$~GeV and $W>3.2$~GeV. We do not find
any evidence of such a component, and no correction
is applied for this effect in the other energies.
We assign a 2\% error from this source for the entire $W$ region.

\item {\it Luminosity function}: We assign 4\% (5\%) for $W$ below (above) 
3.0~GeV; this includes the uncertainties in
the equivalent photon approximation (3\% (4\%)), the
radiative corrections that were neglected (1-2\%) and
the integrated luminosity (1.4\%).

\item {\it No unfolding}: 

Uncertainty from 
smearing effects
is estimated by 
smearing a modeled resonance function with the $W$ resolution
and examining apparent changes of the cross section.
The changes are large ($\sim 7\%$) only near the slopes of the narrowest
resonant structure, in the region $1.44~\GeV \leq W \leq 1.60~\GeV$, 
and smaller (4\%) in other $W$ ranges.

\item {\it Other efficiency errors}: An error of 4\% is assigned for
uncertainties in the efficiency determination based on MC including
the smoothing procedure.
\end{list}

 The total systematic error is obtained by adding all the sources in quadrature
and is 11--12\%
for the intermediate and high $W$ regions. 
It becomes more than 20\% for $W < 1.24~\GeV$.

In the resonance analyses for $W<2.0$~GeV in Sec.~V, we treat the 
systematic error sources except for (9) as uncertainties in the 
overall normalization, which are correlated in the different 
($W$, $|\cos \theta^*|$) bins.  For the 
analysis of the $W$ dependence in the high energy region (Sec.~VI.B), 
we also take into account energy-dependent deviations 
for sources (6) and (8).

\begin{table*}
\caption{Systematic errors for the differential cross sections.
Ranges of errors are shown when they depend on $W$.
}
\label{tab:sysdcs}
\begin{center}
\begin{tabular}{lc} \hline \hline
Source & Error (\%) \\ \hline
Trigger efficiency & 4\\
$\eta$-pair reconstruction efficiency & 6 \\
Overlapping hits from beam background etc. 
& 3 -- 4 \\ 
$p_t$-balance cut  & 3 \\
Sideband background subtraction & 2 -- 27 (for $W$>1.2~GeV) \\
 & 28 -- 60 (for $W$<1.2~GeV) \\
$p_t$-unbalanced background subtraction & 2 \\
Luminosity function and integrated luminosity & 4 -- 5 \\
Unfolding & 4 -- 7 \\

Other efficiency errors & 4 \\ \hline
Overall  & 11 -- 29 (for $W > 1.2~\GeV$)\\
         & 30 -- 61 (for $W < 1.2~\GeV$)\\
\hline\hline
\end{tabular}
\end{center}
\end{table*}

\section{Study of resonances}
\label{sec-5}
In the total cross section (Fig.~\ref{fig07}),
clear peaks due to the $f_2(1270)$
and $f_2'(1525)$ are visible along with other possible resonances.
In this section, 
we first present consistency checks
with previous measurements
and report improved measurements
of some of these resonances.

\subsection{Differential Cross Sections in Partial Waves}
In the energy region $W \leq 3~\GeV$, $J > 4$ partial waves (next is 
$J=6$) may be neglected so that only S, D and G waves are considered.
The differential cross section can be expressed as:


\begin{eqnarray}
\frac{d \sigma}{d \Omega} (\gamma \gamma \to \eta \eta)
 &=& \left| S \: Y^0_0 + D_0 \: Y^0_2  + G_0 \: Y^0_4 \right|^2 \nonumber \\
&+& \left| D_2 \: Y^2_2  + G_2 \: Y^2_4 \right|^2,
\label{eqn:diff}
\end{eqnarray}

\noindent
where $D_0$ and $G_0$ ($D_2$ and $G_2$) denote the helicity 0 (2) components
of the D and G waves, 
respectively\footnote{
We denote individual partial
waves by roman letters and parameterized waves by italic.},
and $Y^{\lambda}_J$ are the spherical harmonics in which the helicity
$\lambda$ is quantized along the $\gamma \gamma$ axis.
Since the $|Y^{\lambda}_J|$'s are not 
independent of each other
partial waves cannot be separated from the information 
in the differential cross sections alone.

We rewrite Eq.~(\ref{eqn:diff}) as


\begin{eqnarray}
\frac{d \sigma}{4 \pi d |\cos \theta^*|}(\gamma \gamma \to \eta \eta)
&=& \nonumber \\
\hat{S}^2 |Y^0_0|^2  + \hat{D}_0^2 |Y^0_2|^2 
&+&\hat{D}_2^2  |Y^2_2|^2  \nonumber \\
+ \hat{G}_0^2  |Y^0_4|^2+ \hat{G}_2^2  |Y^2_4|^2&.&
\label{eqn:diff2}
\end{eqnarray}

The amplitudes $\hat{S}^2$, $\hat{D}_0^2$,  $\hat{D}_2^2$,  $\hat{G}_0^2$
and $\hat{G}_2^2$ can be expressed in terms of $S$, $D_0$, $D_2$, $G_0$ and 
$G_2$~\cite{pi0pi0}.
Since the square of 
spherical harmonics are 
independent of each other,
we can fit differential cross sections to obtain 
$\hat{S}^2$, $\hat{D}_0^2$, $\hat{D}_2^2$, $\hat{G}_0^2$
and $\hat{G}_2^2$ in each $W$ bin.
Since $|Y^0_4|^2$ and $|Y^2_4|^2$ are nearly equal for $|\cos \theta^*| < 0.7$
we also fit $\hat{G}_0^2 + \hat{G}_2^2$
and  $\hat{G}_0^2 - \hat{G}_2^2$.
Two types of fits are made: the ``SD'' fit and ``SDG'' fit.
G waves are neglected in the SD fit.

\begin{figure}
 \centering
   {\epsfig{file=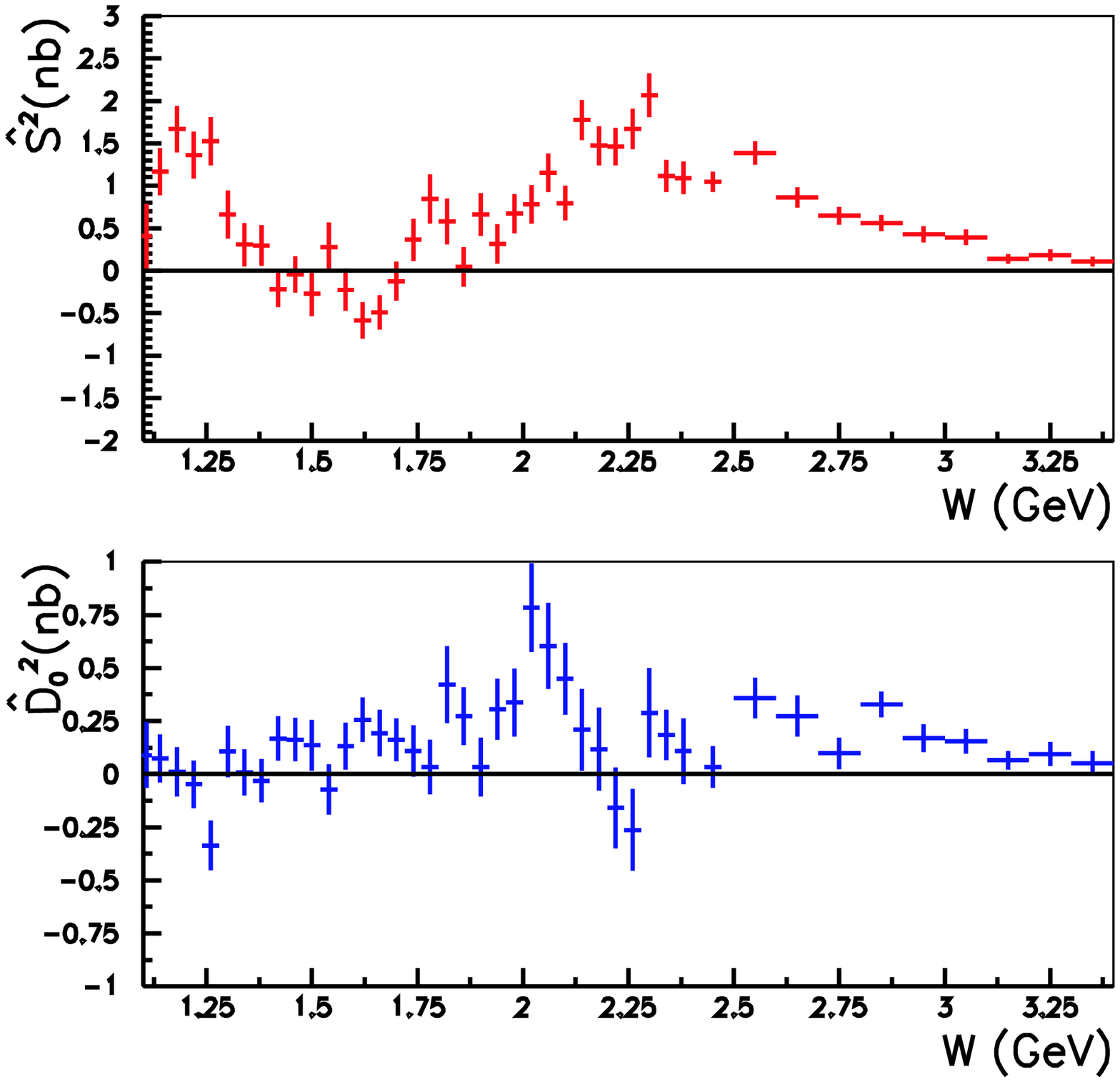,width=50mm}}
   {\epsfig{file=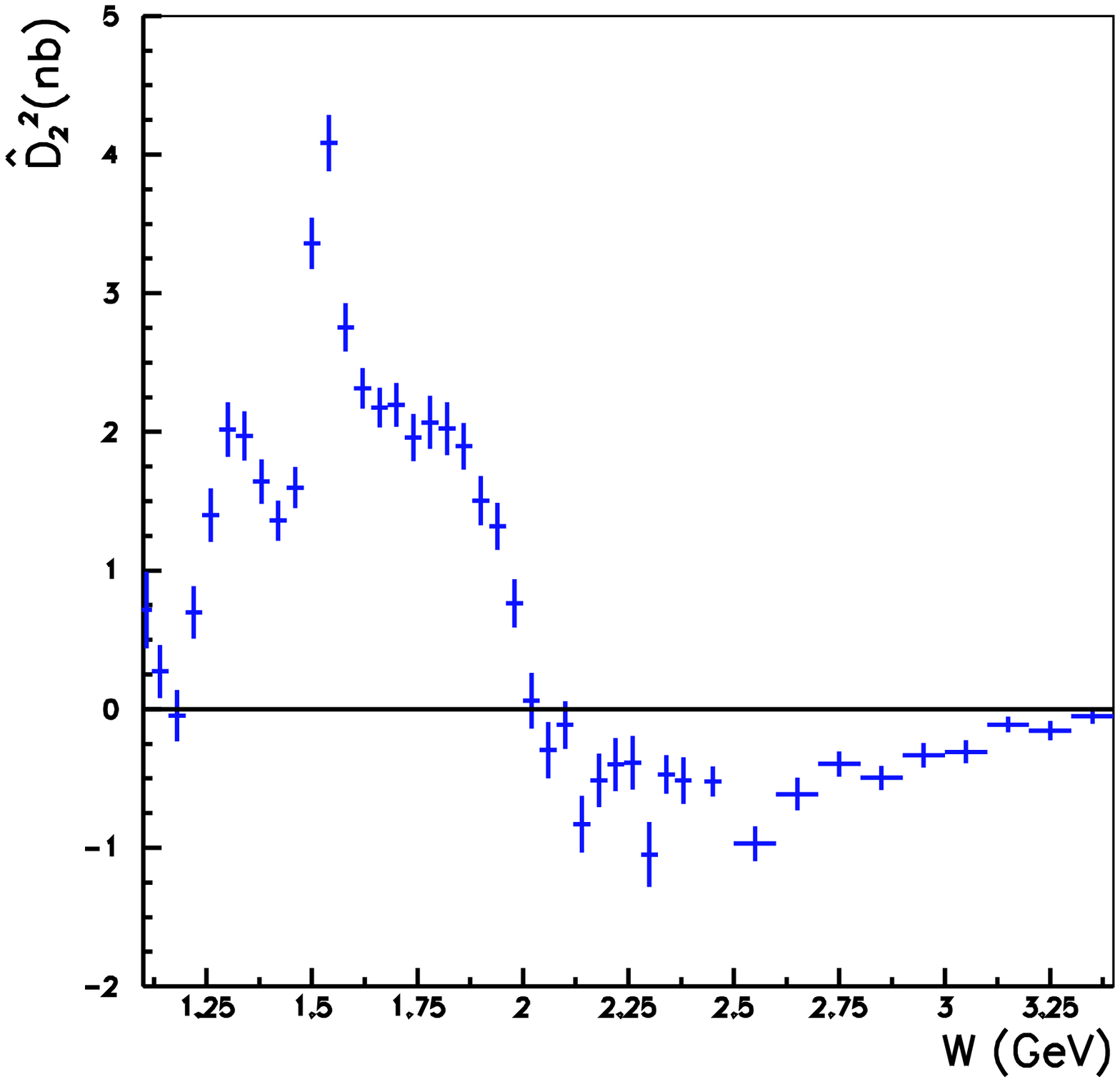,width=50mm}}
 \caption{Spectra of $\hat{S}^2$, $\hat{D}_0^2$ and $\hat{D}_2^2$
for the SD fit. Those for the SDG fit are nearly identical with
larger statistical errors.
The error bars shown are statistical errors
that do not include correlations.
}
\label{fig08}
\end{figure}

\begin{figure}
 \centering
   {\epsfig{file=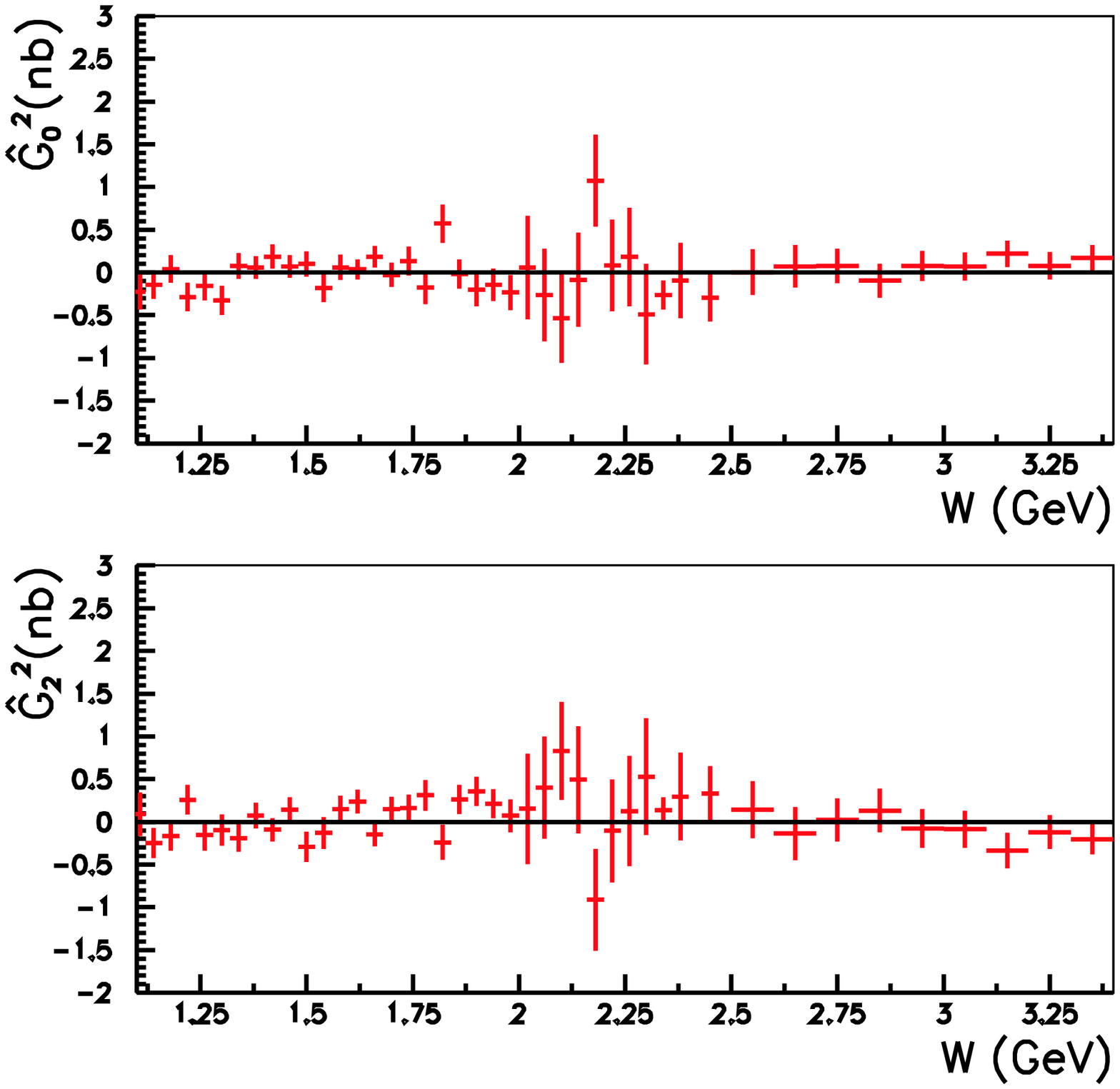,width=50mm}}
   {\epsfig{file=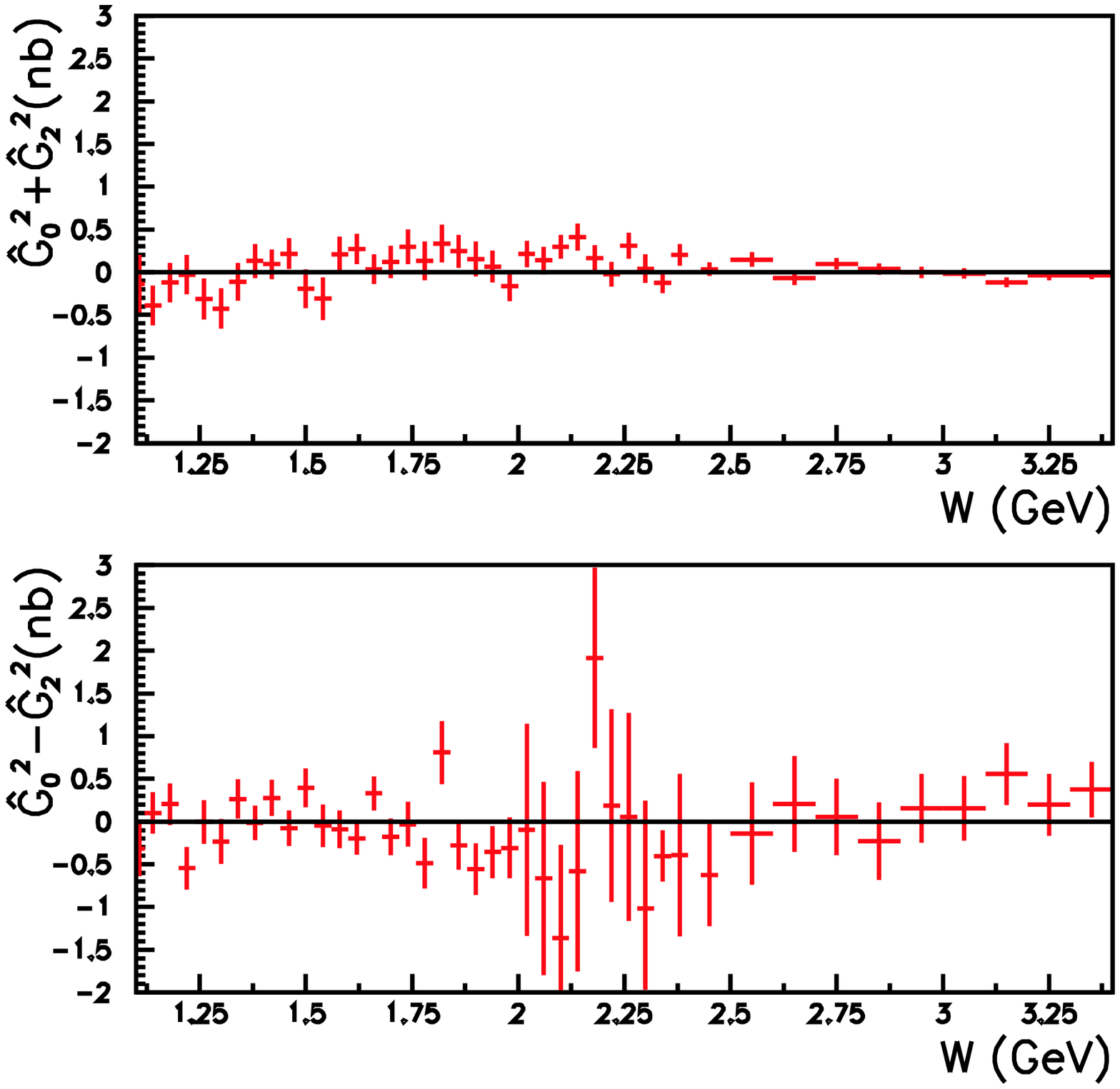,width=50mm}}
 \caption{Spectra of $\hat{G}_0^2$ and $\hat{G}_2^2$ and
$\hat{G}_0^2 \pm \hat{G}_2^2$.
The error bars shown are statistical errors
that do not include correlations.
}
\label{fig09}
\end{figure}

The spectra of $\hat{S}^2$, $\hat{D}_0^2$ and $\hat{D}_2^2$ obtained
for the SD fit and $\hat{G}_0^2$,
$\hat{G}_2^2$ and $\hat{G}_0^2 \pm \hat{G}_2^2$ for the SDG fit
are shown in Figs.~\ref{fig08} and \ref{fig09}.
The spectra of $\hat{S}^2$, $\hat{D}_0^2$ and $\hat{D}_2^2$ for the SDG
fit are omitted because they are nearly the same as those for the SD fit
with 
somewhat larger
statistical errors. 
It appears that the D$_0$ and G waves are small enough to be neglected in the 
region of interest ($W< 2.0~\GeV)$.
In that case, $\hat{S}^2$ and $\hat{D}_2^2$ become $|S|^2$ and $|D_2|^2$,
respectively, which simplifies the parameterization.
In the fits performed here, we neglect the G waves 
completely, and take $D_0 =0$ in the nominal fit.

\subsection{Fitting Partial Wave Amplitudes}
In this subsection, 
we describe the extraction of 
resonant substructure 
by fitting differential cross sections by parameterizing partial wave amplitudes
in terms of resonances and smooth ``backgrounds''.
Note that we do not fit  
$\hat{S}^2$, $\hat{D}_0^2$ and $\hat{D}_2^2$,
but instead fit the differential cross sections directly.
Once the functional forms of amplitudes are assumed, we can use 
Eq.~(\ref{eqn:diff}) to fit differential cross sections. 
We then do not have to worry about the correlations between
$\hat{S}^2$, $\hat{D}_0^2$ and $\hat{D}_2^2$.
The $\hat{S}^2$, $\hat{D}_0^2$ and $\hat{D}_2^2$ spectra are compared 
with the results of parameterization.
Here we neglect the D$_0$ and G waves in the fitting region, $W < 2.00~\GeV$.

Quite a few resonances are listed in Ref.~\cite{PDG} (PDG) that are known 
to decay into $\eta \eta$ with measured or unknown branching fractions
to two photons.
Besides the $f_2(1270)$ and $f_2'(1525)$, there are
$f_2(1565)$, $f_2(1910)$ and $f_2(1950)$  tensor mesons,
$f_0(1370)$, $f_0(1500)$, $f_0(1710)$ and $f_0(2020)$
scalar mesons, and spin-4 $f_4(2050)$ states. So far 
quantitative measurements of the branching fraction to $\eta \eta$ 
based on observed 
enhancements in mass spectra are available for 
$f_2(1270)$~\cite{gams,wa102}
and $f'_2(1525)$~\cite{gams2}. In addition,
a phenomenological derivation of the 
$\eta \eta$ branching fraction 
based on a K-matrix approach~\cite{kmat}
has been tried for the $f'_2(1525)$~\cite{PDG}. 

To investigate this complicated region, we divide our analysis into two parts.
First, we try to confirm or improve the parameter 
$\Gamma_{\gamma \gamma} \B(\eta \eta)$ of the well established
tensor mesons, $f_2(1270)$ and $f_2'(1525)$, by fitting in 
the region $W < 1.64$~GeV.
We then investigate the higher mass region by fixing most of the 
parameters in the fit 
from results in
the low mass region.

\subsubsection{Low mass region, 1.12 -- 1.64~GeV}
We concentrate on the resonances, $f_2(1270)$ and $f_2'(1525)$ by fitting
the region $W<1.64~\GeV$.
%
The resonances taken into account are the $f_2(1270)$,
$f_2'(1525)$ and ``$f_0(Y)$'', where ``$f_0(Y)$'' is just 
a parameterization motivated by the $f_0(1370)$ and $f_0(1500)$.
We parameterize partial waves as follows:
\begin{eqnarray}
S &=& A_{f_0(Y)} e^{i \phi_{Y}} + B_S e^{i \phi_{s}} ,
 \nonumber \\
D_0 &=&  B_{D0}, \nonumber \\
D_2 &=& A_{f_2(1270)} e^{i \phi_{2}} 
+ A_{f_2'(1525)} e^{i \phi_{5}} + B_{D2}, \nonumber \\
\label{eqn:param2}
\end{eqnarray}
where $A_{f_0(Y)}$, $A_{f_2(1270)}$ and $A_{f_2'(1525)}$ 
are the amplitudes of the corresponding resonances;
$B_S$, $B_{D0}$ and $B_{D2}$ are ``background'' amplitudes for 
$S$, $D_0$ and $D_2$ waves; 
$\phi_{Y}$, $\phi_{2}$ and $\phi_{5}$
are the phases of resonances relative to background amplitudes;
$\phi_{s}$ is the relative phase between $S$ and $D_0$.
We set $B_{D0}=0$ (and then $\phi_s = 0$) for simplicity in the
nominal fit, but we later consider a non-zero $D_0$ contribution 
to determine the systematic errors for the 
obtained resonance parameters and leave the $B_{D0}$ symbol
here.

To parameterize resonances,
we use a relativistic Breit-Wigner amplitude
$A_R(W)$ for 
each
spin-$J$ resonance $R$ of mass $m_R$ given by

\begin{eqnarray}
A_R^J(W) &=& \sqrt{\frac{8 \pi (2J+1) m_R}{W}} 
\nonumber \\
&\times&
  \frac{\sqrt{ \Gamma_{\rm tot} (W) \Gamma_{\gamma \gamma}(W) 
{\cal B}(R \rightarrow \eta \eta)}} 
{m_R^2 - W^2 - i m_R \Gamma_{\rm tot}(W)} \; .\nonumber \\
\label{eqn:arj}
\end{eqnarray}

For scalar mesons, partial and total widths do not depend on $W$,
while for tensor mesons (the $f_2(1270)$ and $f_2'(1525)$,
$f_2(1810)$  and $f_2(1950)$),
the energy-dependent total width $\Gamma_{\rm tot}(W)$ is given by
\begin{equation}
\Gamma_{\rm tot}(W) = \sum_X \Gamma_{X \bar{X}} (W) \; ,
\label{eqn:gamma}
\end{equation}
where $X$ is a $\pi$, $K$, $\eta$, $\gamma$, etc.
The partial width $\Gamma_{X \bar{X}}(W)$ is 
parameterized as~\cite{blat}:
\begin{eqnarray}
\Gamma_{X \bar{X}} (W) &=& \Gamma_R {\cal B}(R \rightarrow X \bar{X}) 
\left( \frac{q_X(W^2)}{q_X(m_R^2)} \right)^5 
\nonumber \\
&\times& 
\frac{D_2\left( q_X(W^2) r_R \right)}{D_2 \left( q_X(m_R^2) r_R \right)} \;,
\label{eqn:gamx}
\end{eqnarray}
where $\Gamma_R$ is the total width at the resonance mass,
$q_X(W^2) = \sqrt{W^2/4 - m_X^2}$, $D_2(x) = 1/(9 + 3 x^2 +x^4)$,
and $r_R$ is an effective interaction radius that varies from 1~$\GeV^{-1}$ 
to 7~$\GeV^{-1}$ in different hadronic reactions~\cite{grayer}.
We assume the same $r_R$ value from Ref.~\cite{mori2} 
for the $f_2(1270)$ and $f'_2(1525)$.

For the $4 \pi$ and other decay modes,
$\Gamma_{4 \pi} (W) = \Gamma_R {\cal B}(R \rightarrow 4 \pi)
\frac{W^2}{m_R^2}$ is used instead of Eq.~(\ref{eqn:gamx}) for the $f_2(1270)$.
Parameters of the $f_2(1270)$ and $f_2'(1525)$ 
are summarized in Table~\ref{tab:f2fit}.
The resonance parameters given in Ref.~\cite{PDG} for the 
$f_0(1370)$ 
and $f_0(1500)$ are summarized in Table~\ref{tab:param}.
Background amplitudes are parameterized as follows.
\begin{eqnarray}
B_S &=&  \beta (b_S (W - W_0) + c_S)
, \nonumber \\
B_{D0} &=& \beta^5 (b_0 (W - W_0) + c_0) , \nonumber \\
B_{D2} &=& \beta^5 (b_2 (W - W_0) + c_2) , \nonumber 
\label{eqn:para4}
\end{eqnarray}
where $\beta$ is the velocity of the $\eta$ meson in the c.m.s. and
$W_0 = 2m_\eta$.
We set $B_{D0}=0$, that is, $b_0 = c_0 = 0$, in
the nominal fit.
We assume the background amplitudes for $S$ and $D_2$
to be real and linear in $W$ to reduce 
the number of parameters.
Furthermore, we fix arbitrary phases by choosing 
$c_S > 0$, and $c_2 > 0$.

We fit the energy region of $1.12~\GeV < W < 1.64~\GeV$.
In the fit, we fix the values of the parameters of the $f_2(1270)$ and
$f_2'(1525)$ to those in the PDG~\cite{PDG} except for 
the product
$\Gamma_{\gamma\gamma} \B(\eta\eta)$ for the $f_2(1270)$.

\begin{figure*}
 \centering
   {\epsfig{file=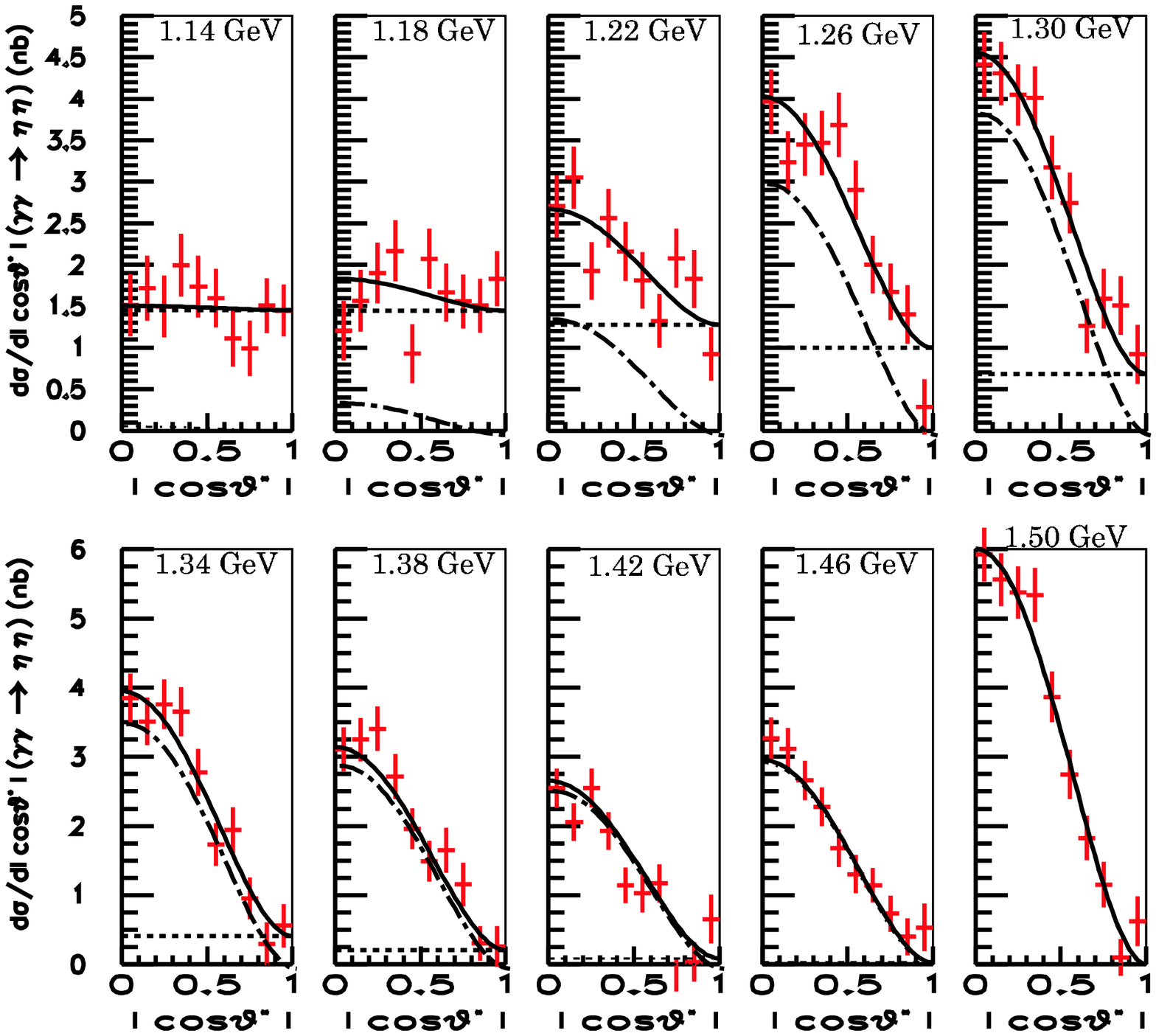,width=100mm}}
   {\epsfig{file=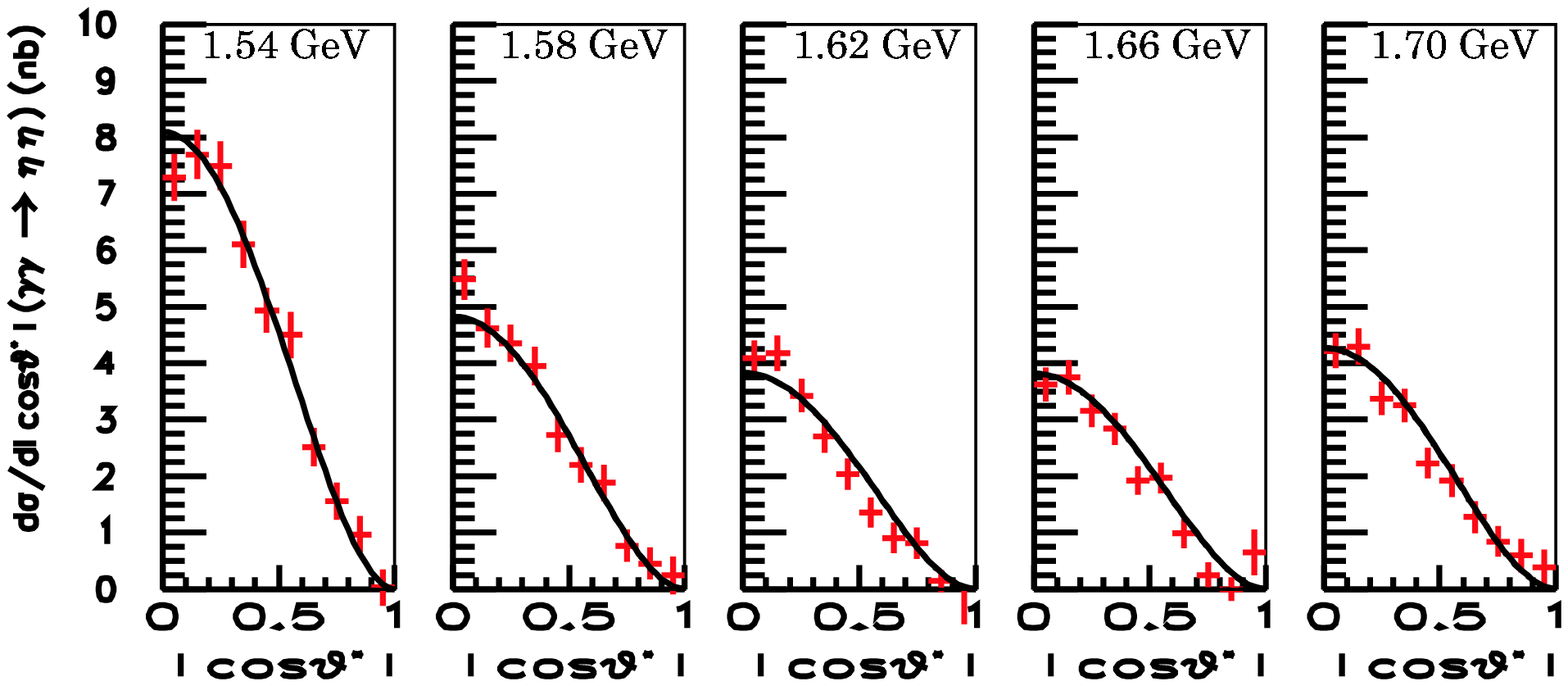,width=100mm}}
 \caption{Differential cross sections 
(points with error bars)
and fitted curves for the nominal fit in the low mass region (solid curve).
Dotted (dot-dashed) curves are $|S|^2$ ($|D_2|^2$) from the fit.
The vertical error bars are statistical only.
}
\label{fig10}
\end{figure*}

\begin{figure}
 \centering
   {\epsfig{file=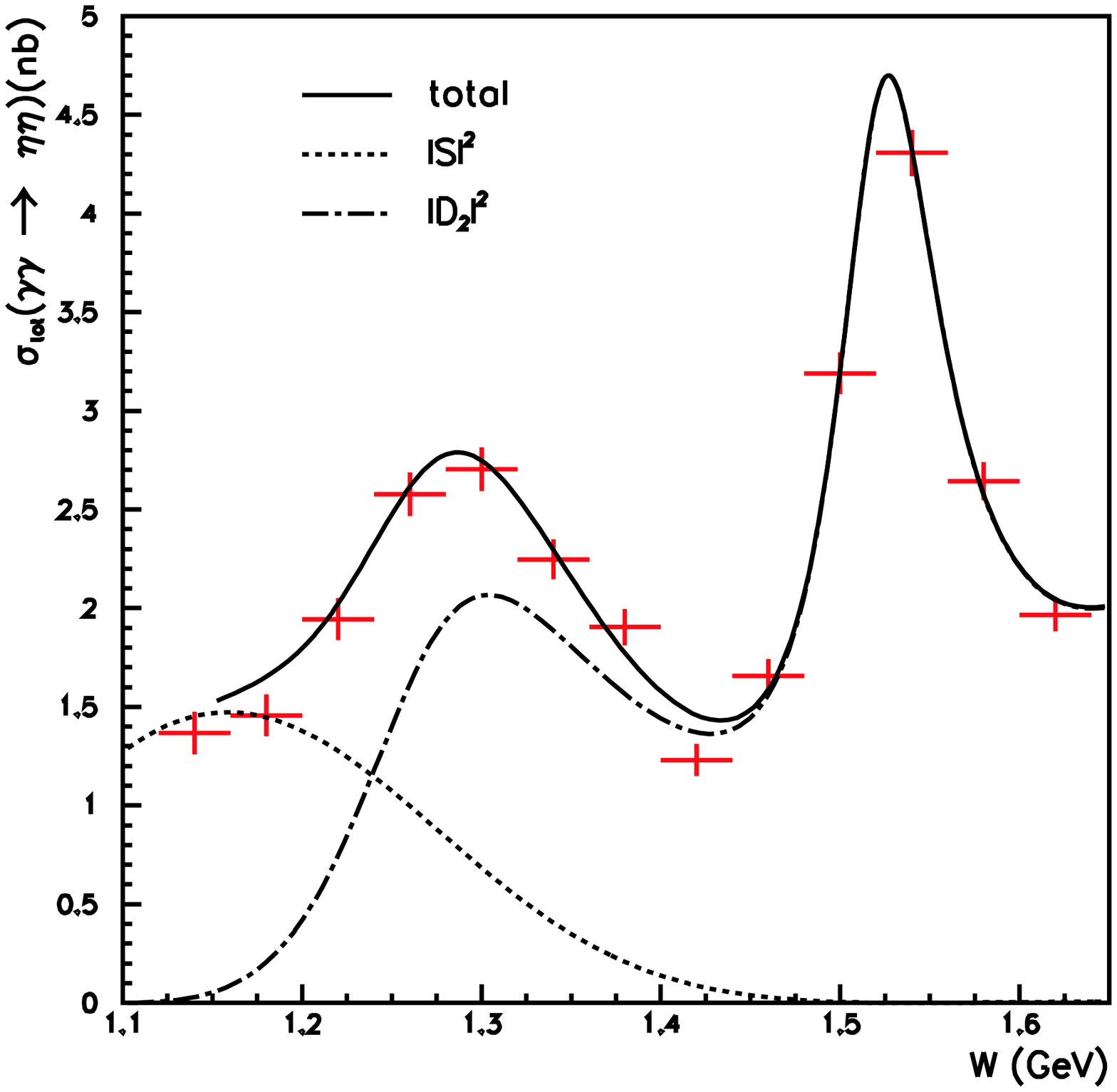,width=50mm}}
 \caption{Total cross section
(points with error bars) 
($|\cos \theta^*| < 1.0 $) and fitted curves
 for the nominal fit in the low mass region (solid curve).
Dotted (dot-dashed) curves are $|S|^2$ ($|D_2|^2$) from the fit.
The vertical errors are statistical only.
}
\label{fig11}
\end{figure}

\begin{figure}
 \centering
   {\epsfig{file=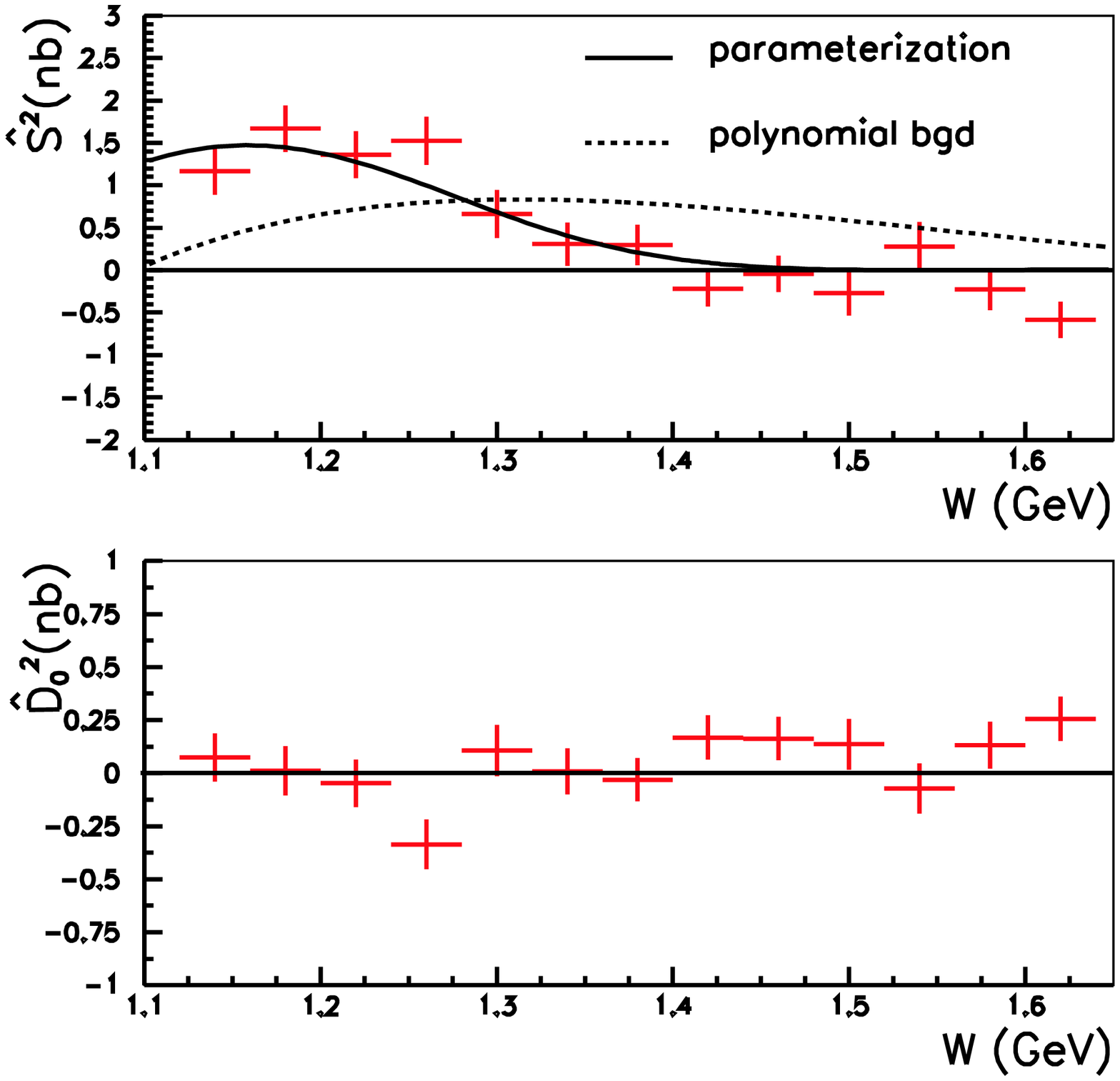,width=50mm}}
   {\epsfig{file=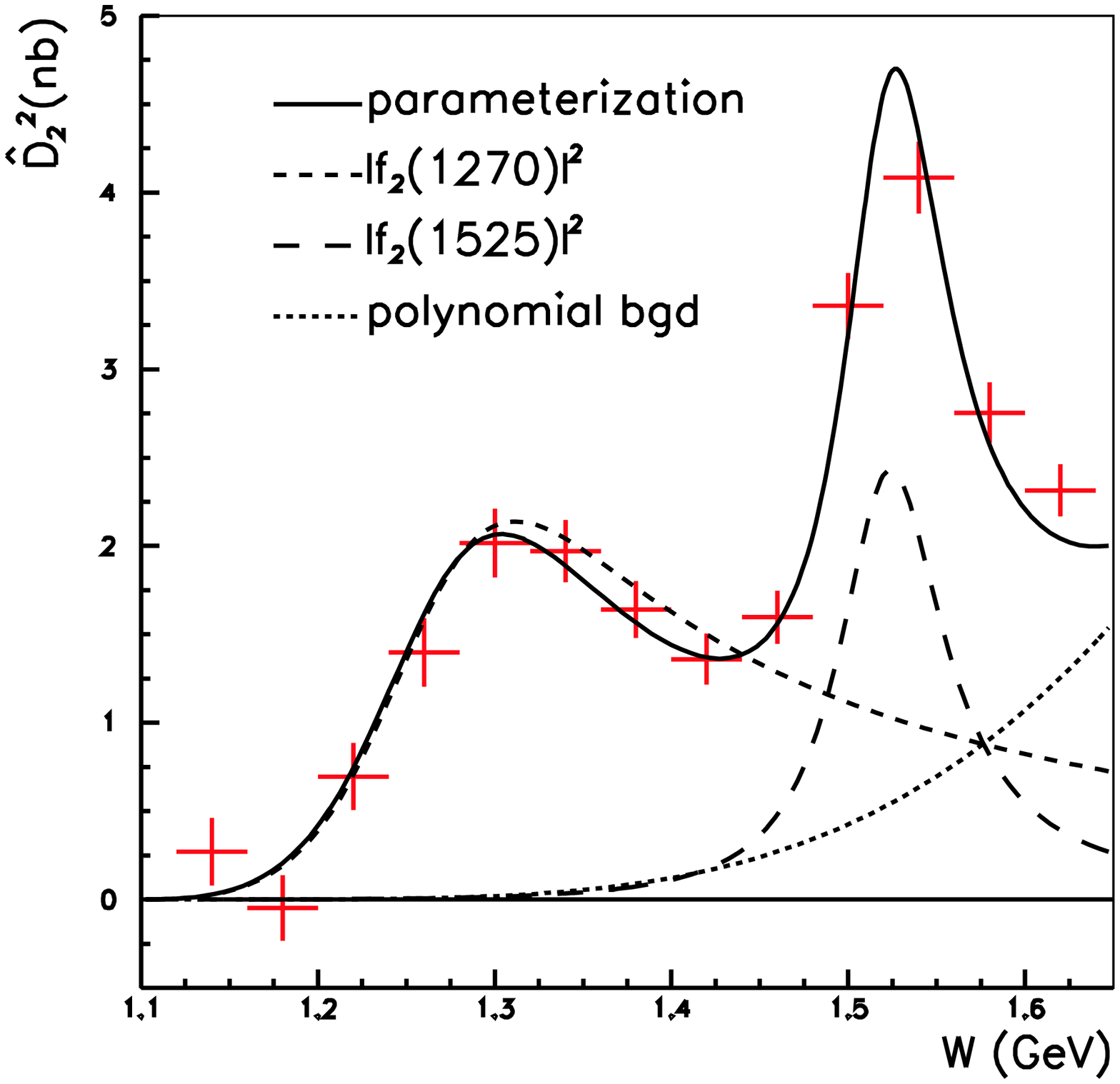,width=50mm}}
 \caption{$\hat{S}^2$, $\hat{D}_0^2$ and $\hat{D}_2^2$
and fitted curves for the nominal fit in the low mass region (solid curve).
The points with error bars are the results
of the $W$-independent fits (same as those in Fig.~8).
The vertical error bars are statistical only.
}
\label{fig12}
\end{figure}

Two hundred sets of randomly generated initial parameters are 
prepared and fits are performed
for each study.
A unique solution is obtained with a fit quality of 
$\chi^2/ndf = 137.1/119$, where $ndf$ is the number of
degrees of freedom in
the fit.
A fit without $f_0(Y)$ gives a poor fit with $\chi^2/ndf = 209.7/123$.
The parameters obtained from these two fits are summarized in 
Table~\ref{tab:fit1}.
The product $\Gamma_{\gamma\gamma} \B(\eta\eta)$ for the $f_2(1270)$ is 
$11.5^{+1.8}_{-2.0}$~eV and is consistent with $12.1 \pm 2.8$~eV in 
PDG~\cite{PDG}.
Figures \ref{fig10} to \ref{fig12} show results of the nominal fit
to differential cross sections, the total cross section,
and spectra of $\hat{S}^2$, $\hat{D}_0^2$ and $\hat{D}_2^2$.

Fits where the value of the product $\Gamma_{\gamma\gamma} \B(\eta\eta)$ of the $f_2'(1525)$
is floated while that of the $f_2(1270)$ is 
fixed to the PDG value, yields three solutions listed in Table~\ref{tab:1525}.
Thus we fix the former to the PDG values in further studies.

The following sources of systematic errors on the parameters are 
considered: dependence on the fitted region, normalization errors of the 
differential cross sections,
assumptions on the background amplitudes,
and the measurement errors of the $f_2(1270)$ and $f'_2(1525)$.

For each study, a fit is made allowing all the parameters to float;
the differences of the fitted parameters from the nominal values
are quoted as systematic errors.
Here too, two hundred sets of randomly generated 
initial parameters are prepared
for each study
and fitted to search for the true minimum and for possible multiple solutions.
Unique solutions are found many times.
Once a solution is found,
several more iterations of the fitting procedure 
are made to confirm
the convergence.

The resulting systematic errors are summarized in Table~\ref{tab:syser}.
Two fitting regions are tried: 
one region that is
shifted lower by one bin ($ 1.08~\GeV \leq W \leq 1.60~\GeV$)
and 
another shifted
higher by one bin ($ 1.16~\GeV \leq W \leq 1.68~\GeV$).
Studies on normalization are divided into those from uncertainties of 
the overall normalization and those from distortion of the spectra in either
$|\cos \theta^*|$ or $W$.
For overall normalization errors, fits are made with two sets of values
of differential cross sections
obtained by multiplying  by 
$(1 \pm \sigma_{\epsilon}(W, |\cos \theta^*|))$, 
where
$\sigma_{\epsilon}$ is the relative efficiency error;
they are denoted as ``normalization$\pm$'' 
in the table.
For distortion studies, $\pm 4$\% errors for $|\cos \theta^*|<1$ and 
$\pm 4$\%/GeV for the $W$ dependence
are assigned, 
based on the uncertainty discussed for (9) in Sec.~IV.E.
Differential cross 
sections are 
modified by multiplying by 
$(1 \pm 0.08 |\cos \theta^*| \mp 0.04)$
and  $(1 \pm 0.08(W {\rm [GeV]} - 1.38))$
(denoted as ``bias:$|\cos \theta^*| \pm$'' and ``bias:$W \pm$'', respectively).
For studies of background (BG) amplitudes, 
either $b_i$ or $c_i$ 
is set to zero for $B_S$ and $B_{D2}$,
while either $b_0$ or $c_0$ is floated for $B_{D0}$.
Finally, the parameters of the $f_2(1270)$, $f_2'(1525)$, and the value of
$r_R$ are successively varied by their errors.

The total systematic errors are calculated by adding individual errors in 
quadrature.
As can be seen in Table~\ref{tab:syser},
we obtain

\begin{equation}
\Gamma_{\gamma \gamma} \B (f_2(1270) \to \eta \eta)
= 11.5~^{+1.8}_{-2.0}~^{+4.5}_{-3.7}~\eV,
\end{equation}
which is consistent with previous measurements~\cite{PDG}.
The apparent threshold enhancement in the S wave is fitted in terms of
a scalar meson, $f_0(Y)$ whose mass, width and 
$\Gamma_{\gamma \gamma} \B (\eta \eta)$
are obtained to be 
\begin{eqnarray}
&&M_{f_0(Y)} = 1262^{+51}_{-78}~^{+82}_{-103}~\MeV/c^2 , \\
&&\Gamma_{f_0(Y)} = 484~^{+246}_{-170}~^{+246}_{-263}~\MeV,\\
&&\Gamma_{\gamma \gamma}\B (f_0(Y) \to \eta \eta) 
= 121~^{+133}_{-53}~^{+169}_{-106}~\eV ,\nonumber \\
\  
\end {eqnarray}
respectively.

The mass peak of the $f_0(Y)$ does not coincide with the broad peak 
in the $\hat{S}^2$ spectrum in Fig.~\ref{fig12} due to the 
effects of interference.

\begin{table*}
\caption{Parameters of the $f_2\lr{1270}$ and $f_2'\lr{1525}$ 
assumed or fitted in 
Ref.~\cite{mori2}.}
\begin{center}
\label{tab:f2fit}
\begin{tabular}{ccccc} \hline \hline
Parameter & $f_2\lr{1270}$ & $f_2'\lr{1525}$  & Unit & Reference \\ \hline
 Mass & $1275.1 \pm 1.2$ & $1525 \pm 5$ & $\MeV/c^2$ & \cite{PDG}\\
 Width & $185.1 ^{+2.9}_{-2.4}$ & $73 ^{+6}_{-5}$ 
& MeV & \cite{PDG}\\
${\cal B}({f_2 \rightarrow \pi \pi})$ & $(84.8^{+2.4}_{-1.2})\%$ 
& $(0.82 \pm 0.15)\%$ 
&  & \cite{PDG} \\
${\cal B}({f_2 \rightarrow K \bar{K}})$ & $(4.6 \pm 0.4)\%$ 
& $(88.7 \pm 2.2)\%$ &  &\cite{PDG} \\
${\cal B}({f_2 \rightarrow \eta \eta })$ & $ (4.0 \pm 0.8) \times 10^{-3}$ 
& $(10.4 \pm 2.2)\%$ &  & \cite{PDG} \\
${\cal B}({f_2 \rightarrow \gamma \gamma})$ & $(1.64 \pm 0.19) \times 10^{-5}$
& $(1.11 \pm 0.14) \times 10^{-6}$ &  & \cite{PDG}\\
$r_R$  & $3.62 \pm 0.03$ & $3.62 \pm 0.03$ 
& $(\GeV/c)^{-1}$ & \cite{mori2} \\
\hline \hline
\end{tabular}
\end{center}
\end{table*}

\begin{table}
\caption{Parameters of the $f_0(1370)$ and $f_0(1500)$~\cite{PDG}.}
\begin{center}
\label{tab:param}
\begin{tabular}{lccc} \hline \hline
Parameter  & $f_0(1370)$  & $f_0(1500)$  & Unit \\
\hline
Mass & 1200 -- 1500 & $1505 \pm 6$ & MeV/$c^2$ \\
Width & 150 -- 250 & $109 \pm 7$ & MeV \\
${\cal B} (\eta \eta)$ & seen & $(5.1 \pm 0.9)\%$ &  \\
${\cal B} (\gamma \gamma)$  & unknown & unknown & \\
\hline\hline
\end{tabular}
\end{center}
\end{table}


\begin{center}
\begin{table*}
\caption{Fitted parameters for the nominal fit and 
for a fit without the $f_0(Y)$
in the low mass region.}
{\renewcommand\arraystretch{1.35}
\begin{tabular}{lccc} \hline \hline
Parameter & Nominal & Without $f_0(Y)$ & Unit\\
\hline
Mass $(f_0(Y))$ &  $1262^{+51}_{-78}$ & -- & $\MeV/c^2$\\
Width $(f_0(Y))$ &  $484^{+246}_{-170}$ & -- & $\MeV$\\
$\Gamma_{\gamma \gamma} \B(\eta\eta)(f_0(Y))$ &  $121^{+133}_{-53}$ 
& 0 (fixed) & eV\\
$\phi_Y$ & $38^{+19}_{-30}$  & -- & deg.\\ 
\hline
$\Gamma_{\gamma \gamma} \B(\eta\eta) (f_2(1270))$ &  $11.5^{+1.8}_{-2.0}$ 
& $11.7^{+1.4}_{-1.5}$  & eV\\
$\phi_{2}$ & $68^{+7}_{-5}$  & $66 \pm 4$  & deg.\\ 
$\phi_{5}$ & $150^{+14}_{-12}$  & $164 \pm 13$   & deg.\\ 
\hline
$b_S$ & $-2.9^{+3.2}_{-3.5}$  & $-8.5^{+0.3}_{-0.4}$   & $\sqrt{\rm nb/GeV}$\\ 
$c_S$ & $2.3^{+2.3}_{-1.8}$   & $3.7 \pm 0.1$  & $\sqrt{\rm nb}$\\ 
$b_2$ & $6.5^{+7.6}_{-5.7}$  & $-3.9^{+3.6}_{-4.5}$   & $\sqrt{\rm nb/GeV}$\\ 
$c_2$ & $1.8^{+3.5}_{-4.6}$  & $6.9^{+3.0}_{-2.3}$   &
$\sqrt{\rm nb}$\\ 
\hline
$\chi^2 \; (ndf)$ & 137.1 (119) &  209.7 (123) & -- \\
\hline\hline
\renewcommand{\arraystretch}{1.0}
\label{tab:fit1}
\end{tabular}
}
\end{table*}
\end{center}

\begin{center}
\begin{table*}
\caption{Three solutions obtained when $\Gamma_{\gamma \gamma} \B(\eta\eta) 
(f_2'(1525))$ is floated in fits to the low mass region.} 
\label{tab:1525}
{\renewcommand\arraystretch{1.35}
\begin{tabular}{lcccc} \hline \hline
Parameter & Sol.A & Sol. B & Sol. C & Unit\\
\hline
Mass $(f_0(Y))$ & $1259^{+50}_{-79}$  & $1259^{+54}_{-80}$  
& $1264^{+55}_{-86}$   & $\MeV/c^2$\\
Width $(f_0(Y))$ & $471^{+234}_{-169}$   & $502^{+241}_{-191}$   
& $536^{+261}_{-193}$   & $\MeV$\\
$\Gamma_{\gamma \gamma} \B(\eta\eta)(f_0(Y))$ & $116^{+122}_{-52}$   
& $127^{+139}_{-59}$  & $143^{+162}_{-69}$   & eV\\
$\phi_Y$ & $36^{+20}_{-31}$ & $38^{+19}_{-32}$ & $41^{+18}_{-30}$  
& deg.\\ 
\hline
$\Gamma_{\gamma \gamma} \B(\eta\eta) (f_2'(1525))$ &  $23.1^{+2.6}_{-2.8}$ 
& $8.0^{+2.0}_{-1.5}$ & $5.0^{+5.8}_{-5.0}$ & eV\\
$\phi_{2}$ & $4^{+10}_{-9}$  & $68^{+10}_{-11}$  & $45^{+23}_{-21}$  & deg.\\ 
$\phi_{5}$ & $188^{+17}_{-14}$  & $155^{+10}_{-11}$  & $94 \pm 22$  & deg.\\ 
\hline
$b_S$ & $-2.9^{+3.2}_{-3.7}$  & $-2.9^{+3.2}_{-3.7}$  & $-2.9^{+3.2}_{-3.6}$  
& $\sqrt{\rm nb/GeV}$\\ 
$c_S$ & $ 2.3^{+2.4}_{-1.9}$  & $2.3^{+1.7}_{-1.9}$  & $2.4^{+2.6}_{-1.9}$  
& $\sqrt{\rm nb}$\\ 
$b_2$ & $1.5^{+6.0}_{-4.8}$  & $3.8^{+6.4}_{-1.9}$  & $-12.5^{+2.5}_{-2.4}$  
& $\sqrt{\rm nb/GeV}$\\ 
$c_2$ & $5.2^{+2.8}_{-3.2}$  & $3.2 \pm 1.2$  & $5.8^{+1.1}_{-1.2}$  
&$\sqrt{\rm nb}$\\ 
\hline
$\chi^2 \; (ndf)$ & 136.4 (119) &  137.2 (119) &  138.6 (119) & -- \\
\hline\hline
\end{tabular}
}
\end{table*}
\end{center}

\begin{table*}
\caption{Systematic uncertainties for the fit in the 
low mass region.}
\label{tab:syser}
\begin{center}
{\renewcommand\arraystretch{1.35}
\begin{tabular}{l|lll|l} \hline \hline 
& \multicolumn{3}{c|}{$f_0(Y)$} & $f_2(1270)$ 
\\ \cline{2-5}
Source & Mass & $\Gamma_{\rm tot}$  & $\Gamma_{\gamma \gamma} \B (\eta \eta)$ 
& $\Gamma_{\gamma \gamma} {\cal B}_{\eta \eta}$\\
& (MeV/$c^2$) & ~~~(MeV)~~~ &~~(eV)~~ & (eV) \\ 
\hline
$W$ range & $^{+56.5}_{-8.9}$ & $^{+0.0}_{-17.8}$ & $^{+26.9}_{-17.0}$ 
& $^{+0.4}_{-1.1}$\\
Bias:$W$ & $^{+1.1}_{-1.7}$& $^{+1.2}_{0.0}$& $^{+2.3}_{-2.1}$
& $^{+0.1}_{-0.1}$ \\
Bias:$|\cos \theta^*|$ 
& $^{+0.1}_{-0.5}$ & $^{+0.5}_{-0.8}$ & $^{+0.6}_{-0.9}$ 
& $^{+0.1}_{-0.1}$\\
Normalization& $^{+27.0}_{-48.8}$ & $^{+220.9}_{-199.1}$ & $^{+152.1}_{-87.1}$ 
& $^{+3.0}_{-2.4}$\\
BG:$B_S$  & $^{+0.0}_{-84.9}$ & $^{+0.0}_{-162.9}$ & $^{+0.0}_{-54.3}$ 
& $^{+0.7}_{-0.0}$\\
BG:$D_0$  & $^{+49.6}_{-0.0}$ & $^{+0.0}_{-42.1}$ & $^{+57.3}_{-0.0}$ 
& $^{+0.0}_{-2.0}$\\
BG:$D_2$  & $^{+4.6}_{-10.1}$ & $^{+100.1}_{-26.7}$ & $^{+30.4}_{-6.0}$ 
& $^{+1.9}_{-0.8}$\\
$f_2$ mass & $^{+0.4}_{-0.2}$ & $^{+4.1}_{-2.5}$ & $^{+2.0}_{-1.4}$ 
& $^{+0.3}_{-0.3}$\\
$f_2$ width & $^{+1.1}_{-0.4}$ & $^{+0.4}_{-0.0}$ & $^{+1.3}_{-0.5}$ 
& $^{+0.2}_{-0.2}$\\
$f_2'$ mass& $^{+8.7}_{-24.7}$ & $^{+41.5}_{-0.0}$ & $^{+21.1}_{-20.3}$ 
& $^{+2.2}_{-1.4}$\\
$f_2'$ width & $^{+4.8}_{-5.6}$ & $^{+2.5}_{-8.4}$ & $^{+0.5}_{-2.7}$ 
& $^{+0.3}_{-0.4}$\\
$f_2' \; \Gamma_{\gamma \gamma} \B (\eta \pi^0)$ 
& $^{+13.0}_{-13.7}$ & $^{+11.7}_{-0.0}$ & $^{+13.0}_{-6.6}$ 
& $^{+1.5}_{-0.4}$\\
$r_R$ & $^{+0.2}_{-0.2}$ & $^{+0.6}_{-1.5}$ & $^{+0.2}_{-0.4}$ 
& $^{+0.0}_{-0.0}$\\ \hline
Total & $^{+81.7}_{-103.0}$& $^{+246.4}_{-262.8}$& $^{+169.4}_{-106.4}$
& $^{+4.5}_{-3.7}$\\
\hline  \hline 
\end{tabular}
}
\end{center}
\end{table*}
\subsubsection{Higher mass region, up to 2.0~GeV}
Now we investigate the higher mass region.
We fix most of parameters determined at lower energy, and 
introduce, just for the purpose of parameterization,
a single tensor resonance,
$f_2(X)$, whose 
mass, width and $\Gamma_{\gamma \gamma} \B(\eta \eta)$ are left free 
 and fit the region $1.16~\GeV < W <2.0~\GeV$.
We parameterize partial waves as follows:
\begin{eqnarray}
S &=& A_{f_0(Y)} e^{i \phi_{Y}} + B_S e^{i \phi_{s}} ,
 \nonumber \\
D_0 &=&  B_{D0}, \nonumber \\
D_2 &=& A_{f_2(1270)} e^{i \phi_{2}} 
+ A_{f_2'(1525)} e^{i \phi_{5}}  \nonumber \\
&& + A_{f_2(X)} e^{i \phi_{X}}+ B_{D2}, \nonumber \\
\label{eqn:hparam2}
\end{eqnarray}
where $A_{f_0(Y)}$, $A_{f_2(1270)}$ and $A_{f_2'(1525)}$, 
are fixed at the values that are fitted in the low mass region.
Here too, $B_{D0}$ is set to zero and $B_S$ is fixed at the values found above.
The phases $\phi_{Y}$, $\phi_{2}$ and $\phi_{5}$
are also fixed and $\phi_{s}=0$.
Only the  $b_2$ and $c_2$ parameters of $B_{D2}$
are floated along with the parameters of $f_2(X)$, 
{\it i.e.}, its 
mass, width, $\Gamma_{\gamma \gamma} \B(\eta \eta)$ and $\phi_{X}$.

\begin{figure*}
 \centering
   {\epsfig{file=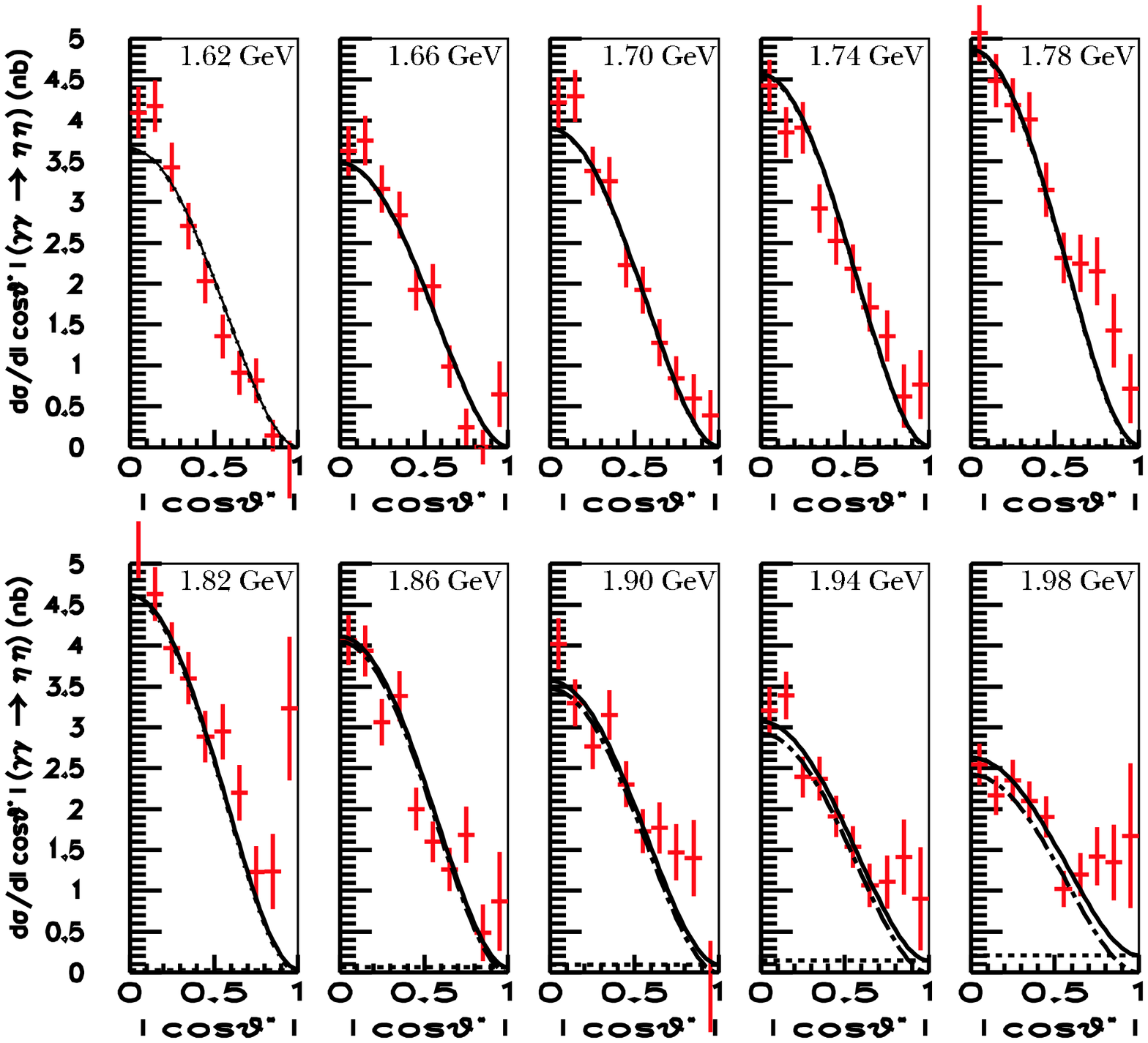,width=100mm}}
 \caption{Differential cross sections 
(points with error bars)
in the energy bins indicated
and fitted curves for the nominal fit in the high mass region (solid curve).
Dotted (dot-dashed) curves are $|S|^2$ ($|D_2|^2$) from the fit.
The vertical error bars are statistical only.
}
\label{fig13}
\end{figure*}

\begin{figure}
 \centering
   {\epsfig{file=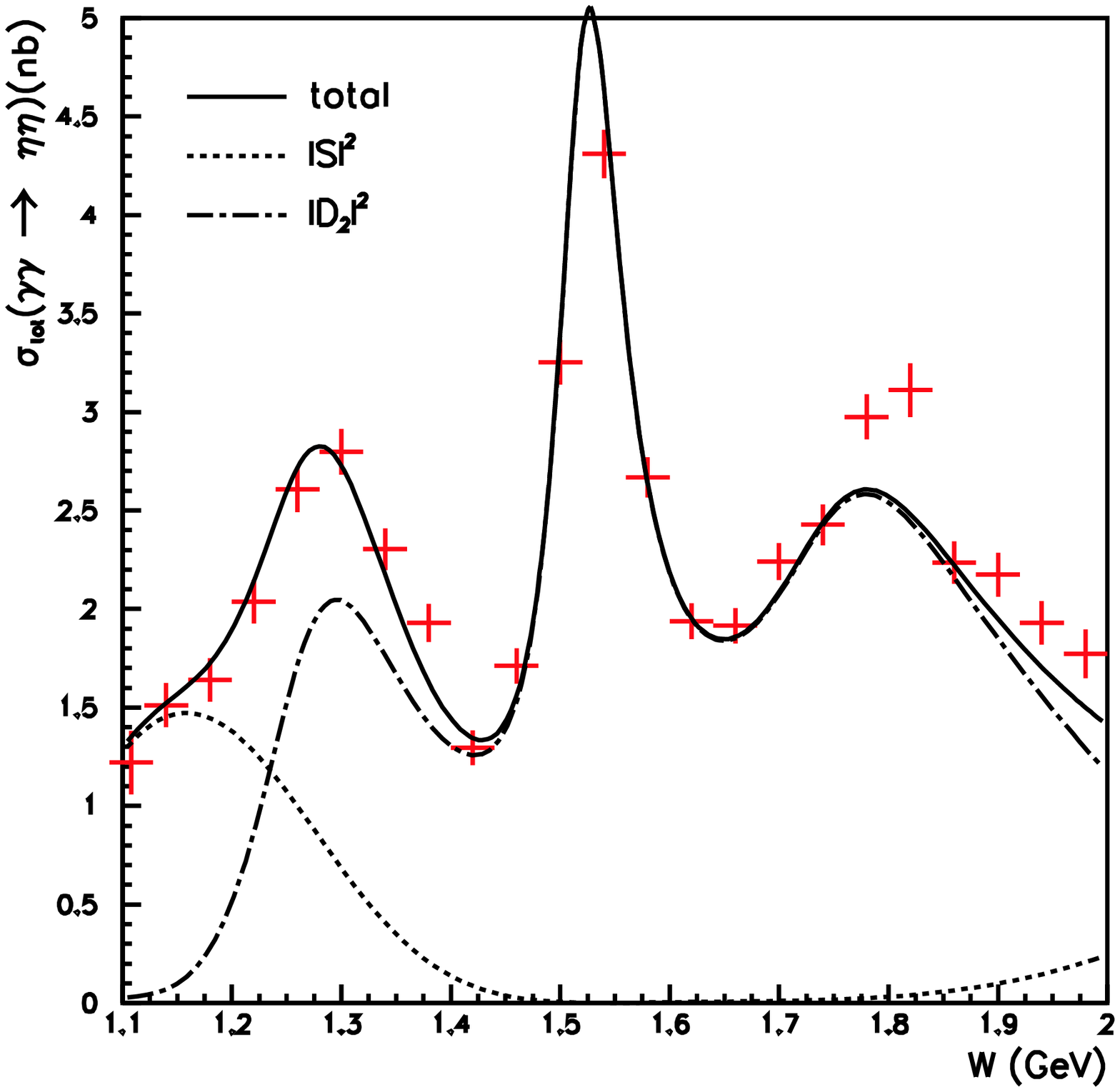,width=50mm}}
 \caption{Total cross sections
(points with error bars)
($|\cos \theta^*| < 1.0 $) and fitted curves for
 the nominal fit in the high mass region (solid curve).
Dotted (dot-dashed) curves are $|S|^2$ ($|D_2|^2$) from the fit.
The vertical error bars are statistical only.
}
\label{fig14}
\end{figure}

\begin{figure}
 \centering
   {\epsfig{file=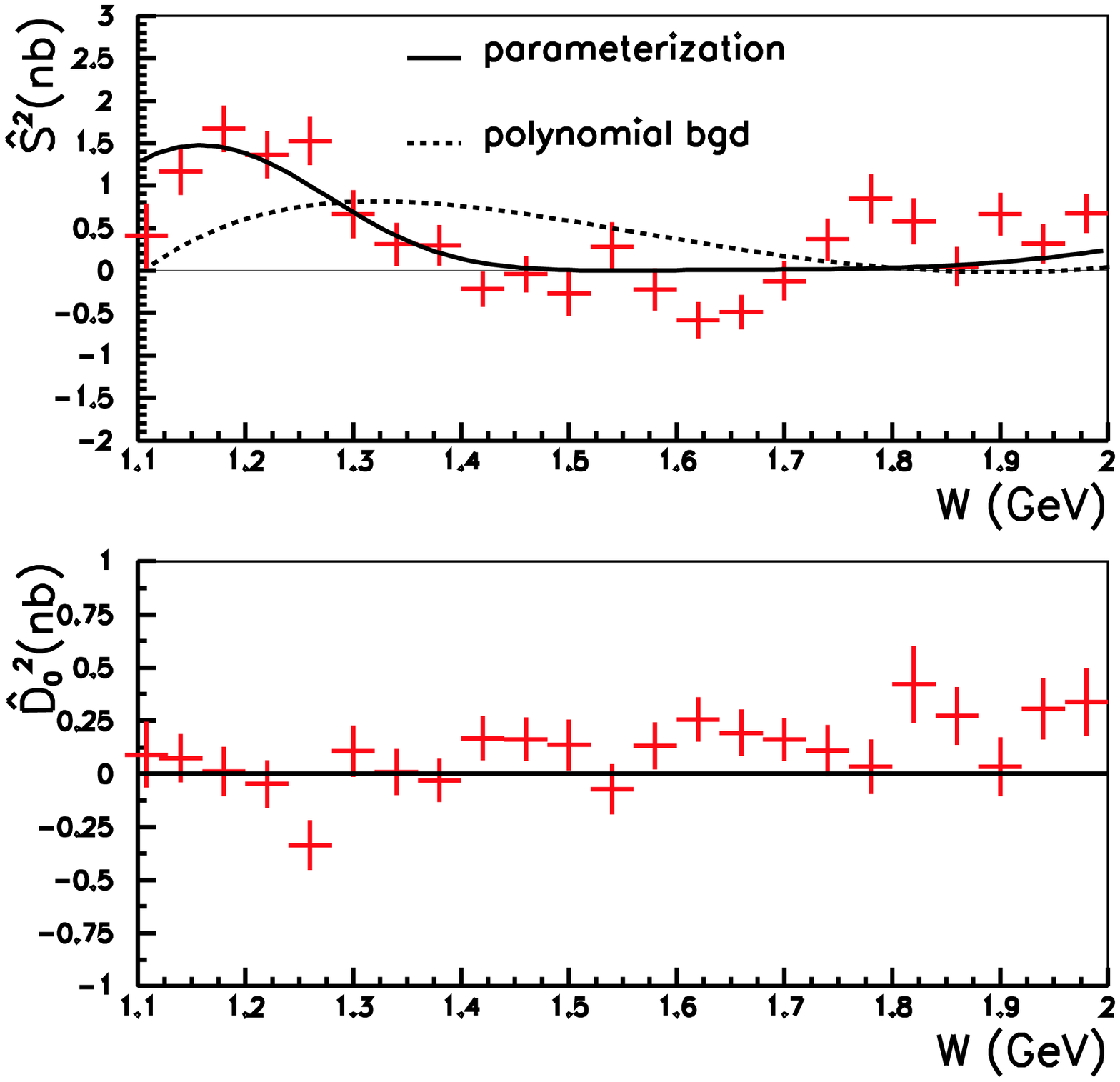,width=50mm}}
   {\epsfig{file=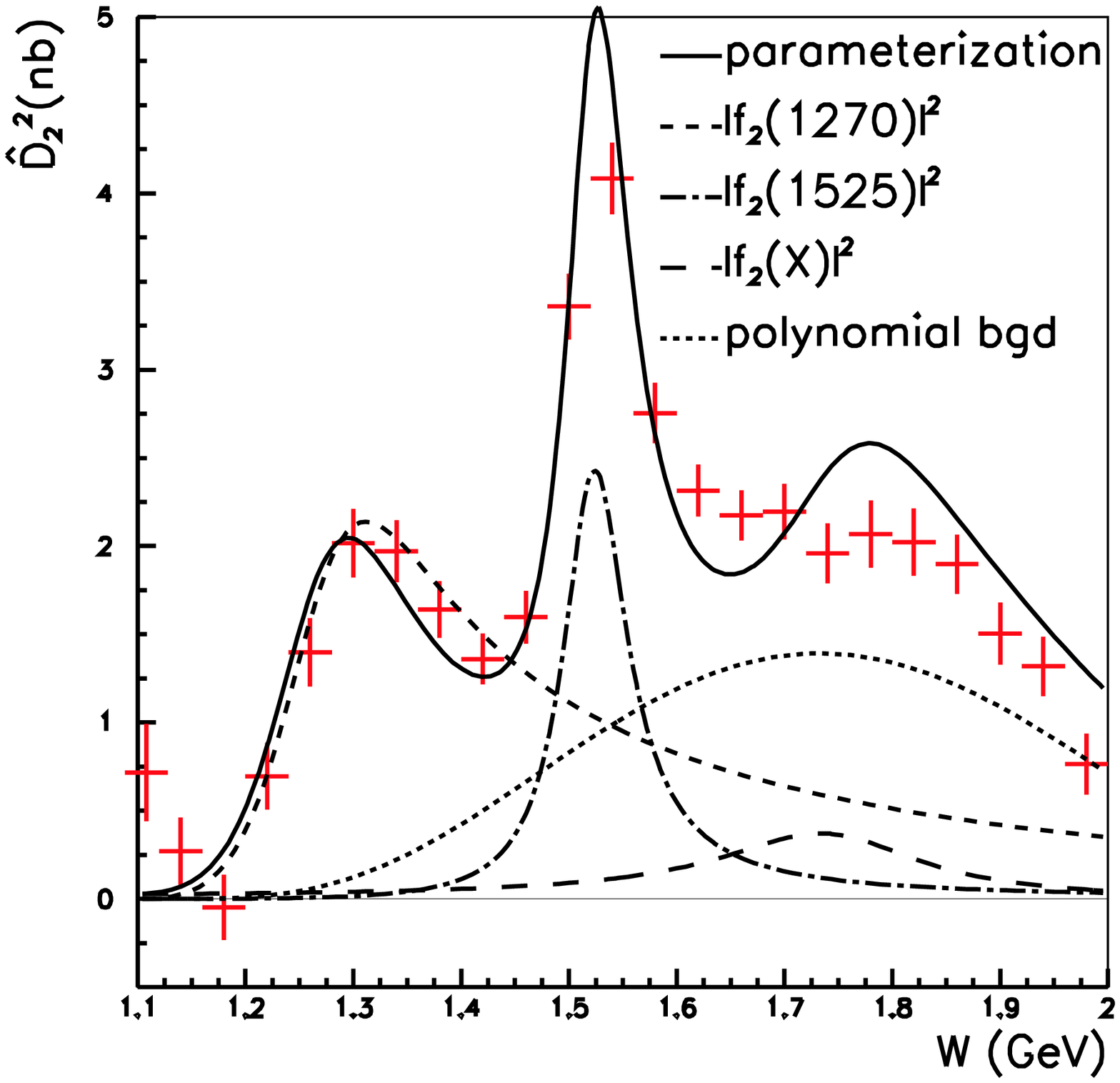,width=50mm}}
 \caption{$\hat{S}^2$, $\hat{D}_0^2$ and $\hat{D}_2^2$
and fitted curves for the nominal fit in the high mass region (solid curve).
The points with error bars are the results of the $W$-independent
fits (same as those in Fig.~8). The vertical error bars are
statistical only.
}
\label{fig15}
\end{figure}

Two hundred sets of randomly generated initial parameters are 
prepared and fits are performed
for each study.
A unique solution is obtained with a fit quality of $\chi^2/ndf = 311.4/204$.
The parameters obtained are summarized in 
Table~\ref{tab:hfit1}.
Figures \ref{fig13} to \ref{fig15} show results of the nominal fit
to the differential cross sections, the total cross section,
and spectra of $\hat{S}^2$, $\hat{D}_0^2$ and $\hat{D}_2^2$.
A more sophisticated parameterization results in multiple solutions.
As an example, two solutions are found when the parameters of $B_S$ are
also floated; these are also listed in Table~\ref{tab:hfit1}.
Hence we employ the simple parameterization given in
Eq.~(13). This parameterization results in 
discrepancies 
from the fits in some $W$ regions 
for differential and integrated 
cross sections. 

Various sources of the systematic errors are studied
and evaluated using various fits similar
to those applied in the analysis for the low mass region,
as summarized in Table~\ref{tab:hsyser}. We take into account
the errors for the $f_0(Y)$ parameters, as well as those for
$f_2(1270)$ and $f'_2(1525)$. We try two fitting regions 
shifted lower by two bins ($ 1.08~\GeV \leq W \leq 1.92~\GeV$)
and higher by two bins ($ 1.24~\GeV \leq W \leq 2.08~\GeV$).
For studies of background (BG) amplitudes, either $c_2$ or $b_2$ is
set to zero for  $B_{D2}$ or allowed to float for $B_{D0}$.
Values of $c_S$ and $b_S$ are changed by their errors.

The total systematic errors are calculated by adding the individual errors in 
quadrature.
The mass, width and 
$\Gamma_{\gamma \gamma} \B (\eta \eta)$ obtained for the $f_2(X)$
meson are 

\begin{eqnarray}
&&M_{f_2(X)} = 1737~\pm 9~^{+198}_{-65}~\MeV/c^2, \\
&&\Gamma_{f_2(X)} = 228~^{+21}_{-20}~^{+234}_{-153}~\MeV,\\
&&\Gamma_{\gamma \gamma} \B (f_2(X) \to \eta \eta)
= 5.2~^{+0.9}_{-0.8}~^{+37.3}_{-4.5}~\eV, \nonumber\\
&&\ 
\end{eqnarray}
 respectively.

The rather poor $\chi^2$ of the fit 
and the clear disagreement in Figs.~\ref{fig14} and
\ref{fig15} above the $f'_2(1525)$
may imply that more than 
one tensor resonance exists in this mass region.
Unfortunately, we cannot draw any definite conclusions 
about such a possibility from additional fits to the data, 
because interference between amplitudes
introduces too much additional freedom.


\begin{table*}
\caption{Fitted parameters for the nominal fit and 
results obtained when the $B_S$ parameters are floated 
for the high mass region.}
\begin{center}
\label{tab:hfit1}
{\renewcommand\arraystretch{1.35}
\begin{tabular}{lc|cc|c} \hline \hline
Parameter & Nominal & \multicolumn{2}{|c|}{Free $B_S$} & Unit\\
\cline{3-4}
&& Sol. A& Sol. B& \\
\hline
Mass $(f_2(X))$ &  $1737 \pm 9$ &  $1742 \pm 10$ &  $1738 \pm 9$ 
& $\MeV/c^2$\\
Width $(f_2(X))$ & $228^{+21}_{-20}$ & $223^{+23}_{-22}$ 
& $236^{+21}_{-20}$  & $\MeV$\\
$\Gamma_{\gamma \gamma} \B(\eta\eta)(f_2(X))$ & $5.2^{+0.9}_{-0.8}$ 
& $4.7^{+1.0}_{-0.9}$ & $6.0^{+1.2}_{-1.0}$  & eV\\
$\phi_{X}$ & $159^{+6}_{-5}$  & $160^{+7}_{-6}$  & $154 \pm 5$  
& deg.\\ 
\hline
$b_S$ & $-2.9$ (fixed)   & $4.4 \pm 0.3$ & $-2.8 \pm 0.3$   
& $\sqrt{\rm nb/GeV}$\\ 
$c_S$ & 2.3 (fixed)  & $0.9 \pm 0.2$  & $1.8 \pm 0.2$  
& $\sqrt{\rm nb}$\\ 
$b_2$ & $-8.1 \pm 0.7$ & $-9.1^{+0.8}_{-0.9}$ & $-9.9 \pm 0.9$   
& $\sqrt{\rm nb/GeV}$\\ 
$c_2$ & $9.4 \pm 0.5$  & $9.8^{+0.6}_{-0.5}$ & $10.4 \pm 0.6$ &
$\sqrt{\rm nb}$\\ 
\hline
$\chi^2 \; (ndf)$ & 311.4 (204) &  279.3 (202) &  288.8 (202) 
& -- \\
\hline\hline
\end{tabular}
}
\end{center}
\end{table*}

{\large
\begin{table}
\caption{Systematic uncertainties for the fit in the high mass region}
\label{tab:hsyser}
\begin{center}
{\renewcommand\arraystretch{1.35}
\begin{tabular}{l|lll} \hline \hline 
& \multicolumn{3}{c}{$f_2(X)$} 
\\ \cline{2-4}
Source & Mass & $\Gamma_{\rm tot}$  & $\Gamma_{\gamma \gamma} \B (\eta \eta)$ 
\\
& (MeV/$c^2$) & ~(MeV)~ &~~(eV)~~  \\ 
\hline
$W$-range & $^{+95.0}_{-0.0}$ & $^{+38.8}_{-101.8}$ & $^{+3.4}_{-1.5}$ \\
Bias:$W$ & $^{+3.8}_{-3.8}$ & $^{+14.2}_{-13.0}$ & $^{+0.6}_{-0.5}$ \\
Bias:$|\cos \theta^*|$ & $^{+0.7}_{-0.8}$ & $^{+6.1}_{-6.1}$ 
& $^{+0.4}_{-0.3}$ \\
Normalization & $^{+125.5}_{-10.4}$ & $^{+142.7}_{-0.0}$ & $^{+16.7}_{-0.0}$ \\
BG:$c_S$ & $^{+0.0}_{-0.0}$ & $^{+0.0}_{-0.0}$ & $^{+0.0}_{-0.0}$ \\
BG:$b_S$ & $^{+0.0}_{-0.0}$ & $^{+0.0}_{-0.0}$ & $^{+0.0}_{-0.0}$ \\
BG:$B_{D0}$ & $^{+0.0}_{-1.4}$ & $^{+0.0}_{-6.2}$ & $^{+0.0}_{-0.5}$ \\
$f_2$ mass & $^{+1.0}_{-1.1}$ & $^{+4.7}_{-4.8}$ & $^{+0.2}_{-0.2}$ \\
$f_2$ width & $^{+1.1}_{-1.4}$ & $^{+5.1}_{-4.1}$ & $^{+0.2}_{-0.2}$ \\
$f_2$ $\Gamma_{\gamma \gamma} {\cal B} (\eta \eta)$ 
& $^{+0.0}_{-8.2}$ & $^{+60.6}_{-0.0}$ & $^{+4.2}_{-0.0}$ \\
$\phi_2$ & $^{+5.1}_{-9.4}$ & $^{+11.8}_{-7.5}$ & $^{+0.4}_{-0.0}$ \\
$f_2'$ mass & $^{+3.8}_{-32.4}$ & $^{+10.4}_{-0.0}$ & $^{+0.6}_{-0.5}$ \\
$f_2'$ width & $^{+6.0}_{-5.9}$ & $^{+7.2}_{-8.1}$ & $^{+0.4}_{-0.4}$ \\
$f_2'$ $\Gamma_{\gamma \gamma} {\cal B} (\eta \eta)$ 
& $^{+30.4}_{-32.4}$ & $^{+4.5}_{-0.0}$ & $^{+0.6}_{-0.0}$ \\
$\phi_5$ & $^{+9.4}_{-13.3}$ & $^{+24.2}_{-15.2}$ & $^{+0.2}_{-0.0}$ \\
$f_0(Y)$ mass & $^{+14.0}_{-21.8}$ & $^{+93.3}_{-41.1}$ & $^{+6.3}_{-1.9}$ \\ 
$f_0(Y)$ width & $^{+36.4}_{-18.2}$ & $^{+70.3}_{-79.7}$ & $^{+3.9}_{-2.7}$ \\ 
$f_0(Y)$ $\Gamma_{\gamma \gamma} {\cal B} (\eta \eta)$ 
& $^{+27.0}_{-20.0}$ & $^{+76.2}_{-66.3}$ & $^{+4.6}_{-2.5}$ \\ 
$f_0(Y)$ phase & $^{+105.7}_{-21.5}$ & $^{+91.8}_{-0.0}$ 
& $^{+31.7}_{-0.0}$ \\ 
$r_R$ & $^{+0.5}_{-0.6}$ & $^{+0.9}_{-0.8}$ & $^{+0.0}_{-0.0}$ \\
\hline
Total & $^{+198.3}_{-65.3}$ & $^{+233.7}_{-153.1}$ & $^{+37.3}_{-4.5}$ \\ 
\hline  \hline 
\end{tabular}
}
\end{center}
\end{table}
}
\section{Analysis of the high energy region above 2.4~GeV}
\label{sec-6}
In this section, we present a study of the angular dependence of the differential cross 
section, the $W$ dependence of the total cross section,
the ratio of cross sections for $\eta \eta$ to $\pi^0 \pi^0$
and $\chi_{cJ}$ charmonium production 
in the high energy region, $W > 2.4~\GeV$.

\subsection{Angular dependence}
As in the analysis of the $\pi^0\pi^0$~\cite{pi0pi02} and 
$\eta \pi^0$~\cite{etapi0} processes, 
we compare the angular dependence of the differential cross sections 
with the function $1/\sin^4{\theta^*}$ for the data in the $W$ range 
2.4~GeV $ < W < 3.3$~GeV. 

In the study of $\pi^0 \pi^0$ data, the contribution from the charmonia is
subtracted~\cite{pi0pi02}.  However, no reliable charmonium subtraction is
possible for the $\eta\eta$ cross section
because of the low statistics and the larger 
charmonium component (see Sec.~IV.D) compared to the $\pi^0\pi^0$
case. We limit our discussion in
Sec.~VI.A-C to the region 
$W<3.3$~GeV 
only, where the 
contribution of charmonium 
is small. 

\begin{figure}
 \centering
   {\epsfig{file=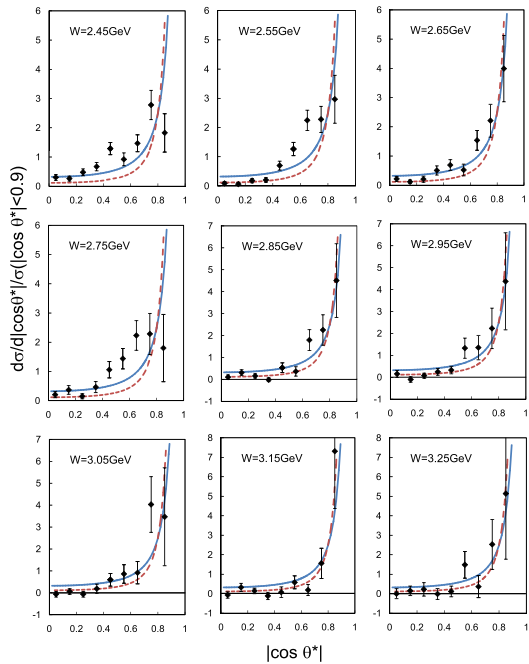,width=70mm}}
 \caption{The angular dependence of the differential cross sections in 
different $W$ regions, with
the normalization to the cross section integrated over $|\cos \theta^*|<0.9$.
The solid and dashed curves are proportional to $1/\sin^4 \theta^*$ 
and $1/\sin^6 \theta^*$, respectively, normalized similarly.}
\label{fig16}
\end{figure}

Figure~\ref{fig16} compares the normalized differential cross sections with
the function, $0.322/\sin^4 \theta^*$ (solid curves). 
The factor in the numerator 
is calculated by dividing differential cross sections, 
which are proportional to $1/\sin^4 \theta^*$
by the total integral for $|\cos \theta^*|<0.9$. 
Agreement is poor in the $W$ region considered.
A $1/\sin^6 \theta^*$ dependence
(dashed curves in the same figure) agrees better 
with the data for $W>3.0$~GeV.
The $\chi^2$'s for the $1/\sin^4 \theta^*$ 
($1/\sin^6 \theta^*$) dependences
are 
29.6 (14.3)
for the $W=3.05$~GeV bin, 
27.8 (7.8) 
for the $W=3.15$~GeV bin, and 
9.8 (4.7) 
for the $W=3.25$~GeV bin.
The number of degrees of freedom is 8, 
and only statistical errors are
used to evaluate the $\chi^2$.

A $1/\sin^4 \theta^*$ dependence is not
a prediction of perturbative QCD (pQCD) for neutral-meson pair 
production, and thus the disagreement does not 
imply an inconsistency with the 
pQCD model~\cite{bl}. 
However, it might indicate that 
the $\eta \eta$ production mechanism is different 
from that of $\pi^0 \pi^0$ and other production
processes where a $1/\sin^4 \theta^*$ 
dependence describes data well for $W>3.1$~GeV.
The handbag model also predicts a $1/\sin^4 \theta^*$ 
dependence for neutral meson pair production processes at 
large Mandelstam variable $t$~\cite{handbag, handbag2}. 
These predictions are critically discussed in Ref.~\cite{newvlc}.
 
\subsection{{\boldmath $W^{-n}$} dependence}
We fit the $W^{-n}$ dependence of the total cross section ($|\cos \theta^*|<0.8$,
where we take the upper boundary 0.8, to match  that
in our $\pi^0\pi^0$ analysis) in the energy region 2.4--3.3~GeV. 
The fit gives
\begin{equation}
n=7.8 \pm 0.6 (stat) \pm 0.4 (sys),
\end{equation}
and the corresponding cross section is shown in 
Fig.~\ref{fig17}(a) together with
that of the $\pi^0\pi^0$ process in the same angular range.  

The systematic error is 
obtained 
by simultaneously varying the 
cross section by $\pm 1 \sigma$ at 2.45~GeV and
$\mp 1 \sigma$ at 3.25~GeV, 
and by $\mp (W~{\rm [GeV]}-2.85)\sigma/0.4$ for
the other $W$ points in between, where
$\sigma$, amounting to 6\%, 
is the systematic error 
that does not include 
the uncertainty in the energy-independent 
normalization.

The slope parameter, $n$, can be compared with $n$ values in 
other processes that we
have studied earlier~\cite{nkzw,wtchen,pi0pi02,etapi0}.
The results are summarized in Table~\ref{tab:val_n}.
The present value for the $\eta\eta$ process is
close to that for the $\pi^0\pi^0$ process, although
we note the measured $W$ regions are different. Differences in
this parameter among different processes are discussed in Ref.~\cite{newvlc}.

\begin{table*}
\caption{The value of $n$ in $\sigma_{\rm tot} \propto W^{-n}$ in
various reactions fitted in the $W$ and $|\cos \theta^*|$ ranges indicated.
The first and second errors are 
statistical and systematic, respectively.
}
\label{tab:val_n}
\begin{center}
\begin{tabular}{lcccc} \hline \hline
Process & $n$ & $W$ range (GeV) & $|\cos \theta^*|$ range & Reference
\\ \hline
$\eta \eta$ & $7.8 \pm 0.6 \pm 0.4$ & 2.4 -- 3.3 & $<0.8$ & This work \\
$\eta \pi^0$ & $10.5 \pm 1.2 \pm 0.5$ & 3.1 -- 4.1 & $<0.8$ & \cite{etapi0} \\
$\pi^0\pi^0$ & $8.0 \pm 0.5 \pm 0.4$ & $ \ \ \ 
$ 3.1 -- 4.1 (3.3 -- 3.6 excluded)
$ \ \ \ $  & $<0.8$ & \cite{pi0pi02} \\
$K^0_S K^0_S$  & $10.5 \pm 0.6 \pm 0.5$ & 2.4 -- 4.0 (3.3 -- 3.6 excluded) 
& $<0.6$ & \cite{wtchen} \\
\hline
$\pi^+\pi^-$ & $7.9 \pm 0.4 \pm 1.5$ & 3.0 -- 4.1 & $<0.6$ & \cite{nkzw} \\
$K^+K^-$  & $7.3 \pm 0.3 \pm 1.5$ & 3.0 -- 4.1 & $<0.6$ & \cite{nkzw} \\
\hline\hline
\end{tabular}
\end{center}
\end{table*}

\subsection{Cross section ratio}
The ratio of cross sections between neutral-pseudoscalar-meson 
($\pi^0$ or $\eta$) pairs in two-photon collisions
can be predicted relatively reliably in both pQCD 
and handbag models, based on quark charges and
flavor-SU(3) symmetry.
The pQCD model~\cite{bl} predictions for the cross section ratios for 
$\pi^0\pi^0$, $\eta\pi^0$
and $\eta\eta$ are summarized in Table~\ref{tabh-1}. 
In the table, $R_f = (f_{\eta}/f_{\pi^0})^2$, where $f_\eta$ ($f_\pi$)
is the $\eta$ ($\pi^0$) form factor.
The value of $R_f$ is not well known, 
and we provisionally assume it to be unity.
The ratio of the cross sections is proportional to the
square of the coherent sum of the product of the quark charges, 
$|\Sigma e_1 e_2|^2$, in which $e_1 = -e_2$ in the
present neutral-meson production cases. 
We show two predictions: a pure flavor-SU(3) octet state 
and a mixture with $\theta_P=-18^\circ$ for the $\eta$ and $\eta'$ mesons.
Here, we assume that the quark-antiquark 
component of the neutral meson wave 
functions dominates and is much larger than
the two-gluon component,
in obtaining the relations between the cross sections.

\begin{figure}
 \centering
   {\epsfig{file=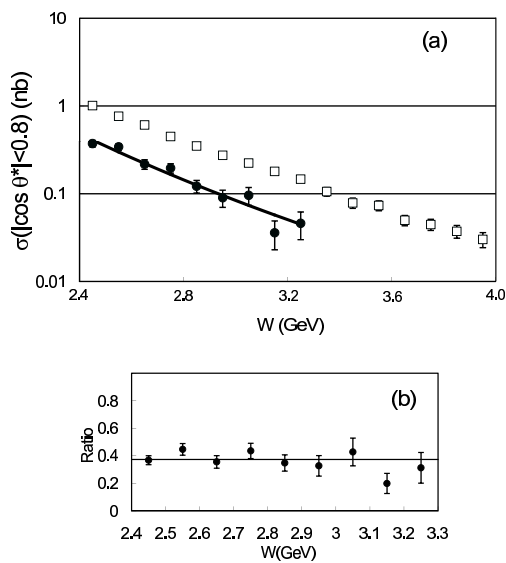,width=70mm}}
 \caption{(a) The $W$ dependence of the cross sections ($|\cos \theta^*|<0.8$)
for the $\pi^0\pi^0$ (open squares)~\cite{pi0pi02}
and $\eta \eta$ (closed 
circles) processes. 
The curve is the power-law fit for the latter process. 
(b) The $W$ dependence of the cross section ratio of $\eta\eta$ 
to $\pi^0\pi^0$ ($|\cos \theta^*|<0.8$).
The line is the average in the 2.4 - 3.3~GeV range.
The error bars are only statistical in the 
above figures.
}
\label{fig17}
\end{figure}

The $W$ dependence of the ratio between the measured cross section 
integrated over $|\cos \theta^*|<0.8$ of 
$\gamma \gamma \to \eta \eta$ to
$\gamma \gamma \to \pi^0 \pi^0$ is plotted in Fig.~\ref{fig17}(b).
For the $\pi^0\pi^0$ process, the
contributions from charmonium production are subtracted using a 
model-dependent assumption described in Ref.~\cite{pi0pi02}.
We use the $\eta \eta$ result only below $W<3.3$~GeV, where 
the charmonium contribution is negligibly small.
Even though the ratio may have a slight $W$ dependence,
in order to compare with QCD (as was done for other processes)
we average the ratio of the
cross sections over the range $2.4~\GeV < W < 3.3~\GeV$ 
and obtain 
\begin{equation}
\frac{\sigma(\eta \eta)}{\sigma(\pi^0 \pi^0)} = 0.37 \pm 0.02 (stat)
\pm 0.03 (sys)
\end{equation}
for $|\cos \theta^*| <0.8$. 
The prediction of this model with $\theta_P = -18^\circ$ and $R_f=1$ 
agrees well with our previous $\eta \pi^0$ measurement~\cite{etapi0}, 
but it is in poor agreement for the $\eta \eta$ process. 
However, we note that the $W$ 
regions are different in the two cases.

The prediction of the $\eta \eta$ cross section for
$|\cos \theta^*|<0.6$ from the handbag model is
presented in Fig.~5 in Ref.~\cite{handbag2}, which
is based on measurements of other meson-pair
production processes. 
We show the results from this measurement, which can be
directly compared with the prediction in Table~\ref{tabh-2}.
Agreement between the measurement and prediction
is fairly 
good.\footnote{We do not give a quantitative 
comparison because Ref.~\cite{handbag2} provides
only a figure without any numerical values.
}
\begin{table*}
\caption{Predictions and data for the cross section 
ratios~\cite{bl} for $\pi^0\pi^0$, $\eta \pi^0$
and $\eta\eta$ production processes 
in two-photon collisions.  
Here, $R_f = (f_{\eta}/f_{\pi^0})^2$, where $f_\eta$ ($f_\pi$)
is the $\eta$ ($\pi^0$) form factor; the value may be taken to be $R_f = 1$.
The $\eta$ meson is treated as a pure SU(3) octet state for the 
entries in the ``octet'' row,
while ``$\theta_P=-18^\circ$'' is the most probable mixing angle between
the octet and singlet states from experiment~\cite{PDG}.
The first and second errors for the data are 
statistical and systematic, respectively.
}
\label{tabh-1}
\begin{center}
\begin{tabular}{c|c|c}
\hline \hline
$\eta$ in SU(3) 
&  
$\sigma(\eta\pi^0)/\sigma(\pi^0\pi^0)$
& 
$\sigma(\eta\eta)/\sigma(\pi^0\pi^0)$ 
\\ 
\hline
Octet & ~~~$0.24 R_f$~~~ & ~~~$0.36 R_f^2$~~~ \\
$\theta_P=-18^\circ$ 
& $0.46 R_f$ & $0.62 R_f^2$ \\ \hline
Data (ref.) & $0.48 \pm 0.05 \pm 0.04$~\cite{etapi0}
& $0.37 \pm 0.02 \pm 0.03$~(this work)\\
($W$ range) & $(3.1~\GeV < W < 4.0~\GeV)$
& $(2.4~\GeV < W < 3.3~\GeV)$ \\
      \hline \hline 
\end{tabular}
\end{center}
\end{table*}

\begin{table}
\caption{Cross section integrated over
$|\cos \theta^*|<0.6$ multiplied by $s^3$.
The first and second errors are 
statistical and systematic, respectively.
}
\label{tabh-2}
\begin{center}
\begin{tabular}{c|c}
\hline \hline
$s$ (GeV$^2$) & $s^3 \sigma (|\cos \theta^*| <0.6)$ (nb GeV$^6$)\\
\hline
6.00 & $38.7 \pm 3.7 \pm 4.3$ \\
6.50 & $33.5 \pm 4.1 \pm 3.6$ \\
7.02 & $28.4 \pm 4.5 \pm 3.1$ \\
7.56 & $38.1 \pm 5.6 \pm 4.3$ \\
8.12 & $17.1 \pm 4.3 \pm 2.1$ \\
8.70 & $21.7 \pm 5.9 \pm 2.6$ \\
9.30 & $18.5 \pm 6.4 \pm 2.4$ \\
9.92 & $11.7 \pm 6.8 \pm 2.0$ \\
10.56 & $21.2 \pm 10.6 \pm 3.5$ 
\\      \hline \hline 
\end{tabular}
\end{center}
\end{table}

\subsection{Extraction of {\boldmath $\chi_{cJ}$} charmonium 
contribution}
\begin{figure}
 \centering
   {\epsfig{file=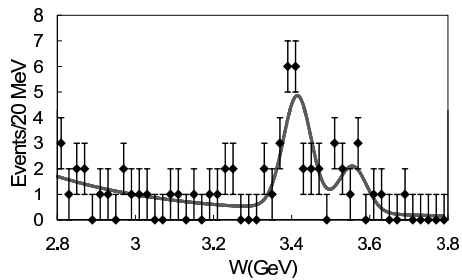,width=60mm}}
 \caption{The $W$ distribution in the charmonium region
($|\cos \theta^*|<0.4$). The fit does not take
interference into account.}
\label{fig18}
\end{figure}

As in our previous $\pi^0\pi^0$ analysis~\cite{pi0pi02}, we extract
the contributions from the $\chi_{c0}$ and $\chi_{c2}$ charmonia
from the $\eta\eta$ data, 
using the raw yield distribution 
in the region 2.8~GeV $<W<3.8$~GeV
integrated over $|\cos \theta^*|<0.4$ (Fig.~\ref{fig18}), 
where the contribution
is enhanced against the forward peak from the QCD effect.

The same formula as in our analysis for the $\pi^0\pi^0$ final 
state~\cite{pi0pi02} is used, where 
partial interference between
the $\chi_{c0}$ charmonium and
the continuum component
is taken into account:

\begin{eqnarray}
&Y&{\hspace{-2mm}}(W)=\nonumber \\
&&|\sqrt{\alpha kW^{-\beta}}+e^{i\phi}\sqrt{N_{\chi_{c0}}}
{\rm BW}_{\chi_{c0}}(W)|^2\nonumber\\ 
&&+N_{\chi_{c2}}|{\rm BW}_{\chi_{c2}}(W)|^2 
+ \alpha (1-k)W^{-\beta}\nonumber,\\
\label{eqn:chicj}
\end{eqnarray}

\noindent
where ${\rm BW}_{\chi_{cJ}}(W)$ is a Breit-Wigner function for the charmonium
amplitude, which is proportional to 
$\sim 1/(W^2-M_{\chi_{cJ}}^2-iM_{\chi_{cJ}} \Gamma_{\chi_{cJ}})$
and is normalized as $\int |{\rm BW}_{\chi_{cJ}}(W)|^2dW=1$.
The masses and widths, $M$ and $\Gamma$, of the charmonium states
are fixed to the PDG world averages~\cite{PDG}.
The component $\alpha W^{-\beta}$ corresponds to the contribution from 
the continuum,
with a fraction $k$ that interferes with the $\chi_{c0}$ amplitude 
with a relative phase angle, $\phi$.  

We do fits with and without interference 
between the $\chi_{c0}$ and the continuum. 
The interference with the $\chi_{c2}$ is neglected because of its narrow width.
We assume a $W$ resolution to be $0.01 W$
from the MC simulation, and take it into account in the fit by smearing
the function $Y(W)$. 
We apply a binned maximum 
likelihood fit with a bin width $\Delta W = 20$~MeV.

The result with interference gives nearly the same result as
the fit without interference but with larger errors.
The fit with interference cannot determine the interference parameters,
$k$ and $\phi$, with a useful accuracy.
Therefore, we take 
the nominal result from the fit without interference.
The best fit is shown in Fig.~\ref{fig18}. 
The results are tabulated in Table~\ref{tabh-3}. 
Significances for the charmonium signals are
5.2$\sigma$ for the $\chi_{c0}$ and 
3.0$\sigma$ 
for the $\chi_{c2}$.
The significances are obtained from the difference of the
logarithmic-likelihoods with and without the
corresponding charmonium contribution, where the change 
in the number of degrees-of-freedom is taken into account. 
Here, in order to obtain the most conservative value, 
we extracted the value
in the interference (non-interference) case for the $\chi_{c0}$ 
($\chi_{c2}$).
The systematic errors are from uncertainties in the $W$ scale and the $W$ 
resolution (we vary
them by $\pm 3$~MeV and by $\pm 20\%$, respectively) 
and the efficiency error. 
 
The results for 
$\Gamma_{\gamma \gamma}(\chi_{cJ}) {\cal B}(\chi_{cJ} \to \eta \eta)$ 
are consistent with the product of the 
known total widths~\cite{PDG} and the branching
fractions from the recent CLEO and BES measurements~\cite{cleobr, besbr},
$(8.0 \pm 0.9)$~eV and $(0.30 \pm 0.04)$~eV for $\chi_{c0}$
and  $\chi_{c2}$, respectively, where we take the average of the
CLEO and BES measurements. 
 
\begin{table*}
\caption{Charmonium yields and $\Gamma_{\gamma\gamma} {\cal B}(\eta \eta)$
from the present measurement. 
Two cases are shown: with and without
interference between $\chi_{c0}$ and continuum.
The first and second (if given) errors are 
statistical and systematic, respectively.
Only differences in log-likelihood
values are meaningful.
}
\label{tabh-3}
\begin{center}
\begin{tabular}{c|c c|c|c c}
\hline \hline
Interference & Yield($\chi_{c0}$) & Yield($\chi_{c2}$) & $-2\ln L/ndf$ & 
$\Gamma_{\gamma\gamma}(\chi_{c0}){\cal B}(\chi_{c0} \to\eta \eta)$~~(eV)&
$\Gamma_{\gamma\gamma}(\chi_{c2}){\cal B}(\chi_{c2} \to\eta \eta)$ \\
\hline
Without & $21.7 \pm 5.3$ & $8.5 \pm 3.6$ & 39.5/46 & 
$9.4 \pm 2.3 \pm 1.2$ & $0.53 \pm 0.22 \pm 0.09$~ \\
With & $21.5 \pm 9.2$ & $10.1 \pm 3.9$ & 38.5/44 & & \\
\hline \hline 
\end{tabular}
\end{center}
\end{table*}

%
%

\section{Summary and Conclusion}
\label{sec-7}
We have measured the cross section of $\gamma \gamma \to \eta\eta$
using a high-statistics data sample
from $e^+e^-$ collisions corresponding to an integrated luminosity
of 393~fb$^{-1}$ with the Belle detector at the KEKB accelerator. 
We obtain results for the differential cross sections
in the center-of-mass energy ($W$) and polar angle ( $|\cos \theta^*|$) 
ranges of 1.096~GeV (the mass threshold)~$ < W < 3.8~\GeV$ and 
up to $|\cos \theta^*| = 0.9$ or 1.0, depending on $W$.

The differential cross sections are fitted in the energy
regions, 1.12~GeV$ < W <$ 1.64~GeV and 1.20~GeV$ < W <$ 2.00~GeV
using a simple parameterization of the S, D$_0$ and D$_2$ waves,
assuming that amplitudes consist of resonances and 
a smooth background.
In the low energy fit, consistency of the parameters of the 
$f_2(1270)$ and $f_2'(1525)$ with previous measurements is checked. 
The apparent threshold enhancement in the S wave is fitted in terms of
a scalar meson, $f_0(Y)$, whose mass, width and 
$\Gamma_{\gamma \gamma} \B (\eta \eta)$
are obtained to be $1262~^{+51}_{-78}~^{+82}_{-103}~\MeV/c^2$,
$484~^{+246}_{-170}~^{+246}_{-263}~\MeV$ and
$121~^{+133}_{-53}~^{+169}_{-106}~\eV$, respectively.
The $f_0(Y)$ is introduced
only to parameterize the data
and may not be a single resonance.

For the energy region of $1.20~\GeV < W < 2.00~\GeV$,
fits are then performed by fixing most of the parameters obtained 
in the low energy region
and by including an additional tensor resonance.
The obtained mass, width and 
$\Gamma_{\gamma \gamma} \B (\eta \eta)$ for the tensor
meson are 
$1737~\pm 9~^{+198}_{-65}~\MeV/c^2$,
$228~^{+21}_{-20}~^{+234}_{-153}~\MeV$ and
$5.2~^{+0.9}_{-0.8}~^{+37.3}_{-4.5}$~eV, 
respectively. 
 The $f_2(X)$ is a parameterization used
to describe the data in 1700~MeV mass
region.
It may represent some of the possible tensor resonances 
in this mass region.

We observe clear signals from $f_2(1270) \to \eta \eta$ and  
$f'_2(1525) \to \eta \eta$ for the first time in two-photon
collisions. The product
$\Gamma_{\gamma \gamma} \B (\eta \eta)$ for the $f_2(1270)$
is $11.5~^{+1.8}_{-2.0}~^{+4.5}_{-3.7}$~eV.
Our $f_2(X)$ may correspond to the $f_2(1810)$ state reported 
in Ref.~\cite{kmat}. 
The result of our measurements for the product
$\Gamma_{\gamma\gamma}{\cal B}(\eta \eta)$ for the $f_2(1270)$ and 
$f'_2(1525)$ are consistent with the 
previously known 
values~\cite{PDG,gams,wa102,gams2,kmat}. 

The angular dependence of the differential cross section
in the 2.4--3.3~GeV region are compared with $\sim 1/\sin^4 \theta^*$
dependence, as found in the $\pi^0\pi^0$ 
process~\cite{pi0pi02} and predicted by the handbag 
model~\cite{handbag,handbag2} for $W > 3.1$~GeV. However,
in the $\eta\eta$ process, a $1/\sin^4 \theta^*$ dependence is not found in the data for the
the energy region where the measurement is performed.


 The slope parameter
$n$
for the cross section,
$\sigma(W) \sim W^{-n}$, in a similar
$W$ region is close to that measured in the $\pi^0\pi^0$ 
process~\cite{pi0pi02}.

The measured cross section ratio, 
$\sigma(\eta \eta)/\sigma(\pi^0 \pi^0) = 0.37 \pm 0.02 \pm 0.03$ 
(for $|\cos \theta^*|<0.8$),
is compared with
the prediction of pQCD~\cite{bl} with 
a pseudoscalar meson mixing angle, 
$\theta_P=-18^\circ$. We find that
the assumption 
for the squared form-factor ratio, $R_f = (f_\eta/f_{\pi^0})^2 = 1$,
which is in a good agreement with the ratio 
$\sigma(\eta \pi^0)/\sigma(\pi^0 \pi^0)$~\cite{etapi0}
cannot reproduce well the $\eta \eta$ measurement.
Our result agrees rather well with the recent handbag model 
prediction~\cite{handbag2}.

Charmonium contributions in the $\eta \eta$ process are confirmed for the first time. 
Our measurements are consistent with the known
partial decay widths of the $\chi_{c0}$ and $\chi_{c2}$ to 
$\gamma\gamma$~\cite{PDG} and $\eta \eta$~\cite{cleobr, besbr} final states.

\section*{Acknowledgments}
%
We thank the KEKB group for the excellent operation of the
accelerator, the KEK cryogenics group for the efficient
operation of the solenoid, and the KEK computer group and
the National Institute of Informatics for valuable computing
and SINET3 network support.  We acknowledge support from
the Ministry of Education, Culture, Sports, Science, and
Technology (MEXT) of Japan, the Japan Society for the 
Promotion of Science (JSPS), and the Tau-Lepton Physics 
Research Center of Nagoya University; 
the Australian Research Council and the Australian 
Department of Industry, Innovation, Science and Research;
the National Natural Science Foundation of China under
contract No.~10575109, 10775142, 10875115 and 10825524; 
the Ministry of Education, Youth and Sports of the Czech 
Republic under contract No.~LA10033;
the Department of Science and Technology of India; 
the BK21 and WCU program of the Ministry Education Science and
Technology, National Research Foundation of Korea,
and NSDC of the Korea Institute of Science and Technology Information;
the Polish Ministry of Science and Higher Education;
the Ministry of Education and Science of the Russian
Federation and the Russian Federal Agency for Atomic Energy;
the Slovenian Research Agency;  the Swiss
National Science Foundation; the National Science Council
and the Ministry of Education of Taiwan; and the U.S.\
Department of Energy.
This work is supported by a Grant-in-Aid from MEXT for 
Science Research in a Priority Area ("New Development of 
Flavor Physics"), and from JSPS for Creative Scientific 
Research ("Evolution of Tau-lepton Physics").

\end{document}